\documentclass[aps,prb,notitlepage,nofootinbib]{revtex4-2}
\usepackage{graphicx}
\usepackage{amsfonts}
\usepackage{amssymb}
\usepackage{amsmath}
\usepackage{mathtools}
\usepackage{afterpage}
\usepackage[mathscr]{euscript}
\usepackage{braket}
\usepackage{subcaption}
\usepackage{hyperref}
\hypersetup{
    colorlinks=true, 
    linkcolor=blue,  
}

\let\savecorresponds\corresponds
\let\corresponds\relax

\usepackage{mathabx}
\let\corresponds\savecorresponds

\usepackage{amsthm}

\usepackage{multirow}
\usepackage{float}
\usepackage{qcircuit}
\usepackage{leftidx}
\usepackage[utf8]{inputenc} 
\usepackage{yfonts}
\usepackage{dsfont}
\usepackage{cancel}
\usepackage{booktabs}
\usepackage[scr=boondox]{mathalpha}
\usepackage{pst-all}

\newcommand{\coho}[1]{\textswab{#1}}

\begin{document}
	\def\U#1{\mathrm{U}(1)}
	\def\SO{\mathrm{SO}}
	\def\SU{\mathrm{SU}}
	\def\Sp{\mathrm{Sp}}
	\def\H{\mathcal{H}}
	\def\F{\mathcal{F}}
	\def\E{\mathbb{E}}
	\def\TT{\mathsf{T}}
	\def\A{\mathcal{A}}
	\def\L{\mathcal{L}}
	\def\w{\mathfrak{w}}
	\def\O{\coho{O}}
	\def\C{\widecheck{\mathcal{C}}}
	
\newcommand{\Aut}{\text{Aut}}
\newcommand{\Z}{\mathbb Z}
\newcommand{\R}{\mathbb R}

\newcommand{\personalcomment}[1]{\textcolor{red}{[#1]}}
	\newcommand{\commentdb}[1]{\textcolor{blue}{[DB: #1]}}
		\newcommand{\commentmb}[1]{\textcolor{red}{[MB: #1]}}

\def\Sq{\mathop{\mathrm{Sq}}\nolimits}

\newcommand{\mathbbm}[1]{\text{\usefont{U}{bbm}{m}{n}#1}}

    \title{Defect Networks for Topological Phases Protected By Modulated Symmetries}
        
    \author{Daniel Bulmash}
    \affiliation{Department of Physics, United States Naval Academy, Annapolis MD 21402 USA}

    \begin{abstract}
        Modulated symmetries are internal symmetries which do not commute with spatial symmetries; dipolar symmetries are a prime example. We give a general recipe for constructing topological phases protected by modulated symmetries via a defect network construction, generalizing the crystalline equivalence principle to modulated symmetries. We demonstrate that modulated symmetries can be treated identically to unmodulated symmetries in the absence of spatial symmetries, but in the presence of spatial symmetries, some defect networks which are non-anomalous for unmodulated symmetries become anomalous for modulated symmetries. We apply this understanding to classify symmetry-protected topological phases protected by translation symmetry plus either discrete or continuous dipolar symmetries in (1+1)D and (2+1)D and obtain a number of other (1+1)D classification results for modulated symmetry-protected topological phases.
    \end{abstract}

    \maketitle
    \tableofcontents
        
    \section{Introduction}
    
    Symmetry has been a unifying tool in physics for understanding physical phenomena and categorizing phases of matter. For example, we have been able to thoroughly understand symmetry-protected topological phases (SPTs)~\cite{Pollman1DSPT,Chen1DSPT,Chen1DSPT2,Schuch1DSPT,ChenCohomology,SenthilSPTReview}, namely, phases of matter which cannot be adiabatically deformed to a trivial state in the presence of a symmetry but can be so deformed if the symmetry is broken, when they are protected by ordinary global symmetries. In recent years, attempts to use symmetry principles to describe novel physical phenomena have required us to expand our definition of symmetry beyond ordinary global symmetries~\cite{GaiottoGeneralizedSymm,McGreevyGeneralizedSymm,GomesHigherForm,SakuraGeneralizedSym,LuoGeneralizedSymm,CostaSimonsLecture} to include higher-form, higher group, non-invertible, categorical, and many other forms of symmetries.
    
    Within this landscape, modulated symmetries, that is, internal symmetries which do not commute with spatial symmetries, have come under increasing study. A well-studied example of a multipolar symmetry is a dipolar symmetry, that is, a symmetry associated with a conserved $\U1$ or $\Z_N$ dipole moment. This symmetry leads to many exotic phenomena, including localization~\cite{PaiDipoleCircuits,HanMultipoleLocalization}, slow dynamics~\cite{FeldmeirSubdiffusion,GromovFractonHydro,JainDipoleHydro}, Hilbert space fragmentation~\cite{SalaDipoleErgodicityBreaking}, charge immobility~\cite{PretkoSubdimensional}, and unusual gauge theories~\cite{GorantlaGlobalDipole,PretkoFractonGauge}. The list of modulated symmetries has grown rapidly to include multipolar~\cite{GromovMultipole}, subsystem~\cite{VijayHaahFu}, and fractal~\cite{CastelnovoGlassiness,DevakulFractal} symmetries, among others~\cite{BulmashGeneralizedGauge}. General modulated symmetries have interesting consequences in dynamics~\cite{SalaModulatedDynamics,LianQuantumBreakdown} and in their topological aspects~\cite{YouSubsystemSPT,DevakulFractalSPT,DevakulPlanarSSPT,han2023modulated,lam2023dipole,LamDipoleInsulator,EbisuMultipolar,PaceGaugingModulated}, but focusing on topology leads to a puzzling question - what is the fundamental distinction between a modulated and an unmodulated symmetry if the two types of symmetry form the same group? In particular, is there a difference between an SPT protected by an ordinary internal symmetry and one protected by a modulated symmetry? Our aim will be to answer this question and to construct machinery that may be used to classify modulated SPTs (MSPTs), that is, SPTs protected by modulated symmetries.
    
    From the perspective of topology, one might assume that a modulated symmetry is just an internal symmetry; nowhere in the definition of internal symmetries do we require any particular structure of the symmetry operator. Even if we require the internal symmetry to act onsite, there no constraint on the onsite action of the symmetry. The interplay of spatial and multipolar symmetries has had some study~\cite{GromovMultipole,BulmashMultipole}, but largely starting from a spatial symmetry and determining the consequences on multipolar symmetries. In this work, we will demonstrate how, at least from the perspective of topological phases, modulated symmetries are distinct from unmodulated symmetries in the presence of spatial symmetries, but can be treated as ordinary internal symmetries in the absence of spatial symmetries. (As such, we do not use the word ``crystalline" in our MSPT terminology because a ``modulated symmetry" already requires some spatial symmetry to be meaningful.)

    If modulated symmetries are only distinct from ordinary internal symmetries in the presence of spatial symmetries, we are forced to discuss crystalline symmetry-protected topological phases (CSPTs), that is, SPTs protected by spatial symmetries (and also possibly internal symmetries). A key idea in understanding CSPTs is the ``crystalline equivalence principle"~\cite{ElseThorngrenCrystallineEquiv}, which states that the classification of topological phases with symmetry group $G$, where $G$ may include spatial symmetries, is the same as the classification of topological phases where $G$ is treated (formally) as a purely internal symmetry. This idea, which has been extended to quasicrystals~\cite{QuasicrystalTopology}, is supported by the defect network (or, for SPTs, the ``block state")~\cite{ElseThorngrenDefectNetworks,HuangBlockStates,SongBlockState} construction, which explicitly constructs crystalline topological phases in real space with symmetry group $G$ from internal SPT data. Beyond their initial use in bosonic CSPTs, defect networks have been used to generate Lieb-Schultz-Mattis theorems~\cite{JiangLuGeneralizedLSM} and to classify interacting fermionic CSPTs~\cite{FermionicCSPTs,FermionicCrystallineTI}. More generally, the defect network construction has proved useful for applying the tools of topological quantum field theory, which are most naturally suited to topological phases with internal symmetries, to systems with spatial symmetries or other nontrivial spatial structures like position-dependent superselection sectors, i.e., fracton topological orders~\cite{AasenDefectNetworks,WenNonLiquid,WangNonLiquid}, and also to construct Floquet codes~\cite{DefectNetworkFloquet}.
    
    Our aim in this work is to generalize the crystalline equivalence principle and corresponding defect network picture to MSPTs and to obtain classifications of MSPTs for several interesting symmetries. Our results clarify the crucial interplay of spatial symmetries and modulated symmetries. Several prior works have classified what we will call ``strong" MSPTs in certain cases; Refs.~\cite{han2023modulated,lam2023dipole,EbisuMultipolar} have done this for $\Z_N$ dipolar or multipolar symmetries in (1+1)D using matrix product state approaches, and Ref.~\cite{PaceModulatedSymTFT} has used the ``SymTFT" approach to, in principle, classify general (1+1)D modulated symmetries. Our framework is extremely general. It applies to arbitrary dimensions and symmetry groups, can be applied equally well to fermionic and bosonic systems, and can be used for invertible or non-invertible topological phases. Classification is most straightforward for low-dimensional bosonic SPTs, but this is not an intrinsic limitation of the construction.

    Before proceeding, we briefly outline the rest of this paper and summarize our results.

    In Sec.~\ref{sec:modulated}, we define and give some examples of modulated and unmodulated symmetries and show that in the absence of spatial symmetries, finitely generated groups of modulated symmetries can be treated like ordinary internal symmetries. In Sec.~\ref{sec:DefectNetworks}, we review the defect network construction of CSPTs and emphasize aspects that are most important for generalizations to MSPTs. In Sec.~\ref{sec:modulatedDefect}, we generalize the defect network construction to modulated symmetries and derive ('t Hooft) anomaly-free conditions on the input data. The anomaly-free conditions are our most important results. In particular, we show that for a $(d+1)$-dimensional MSPT, just like for CSPTs, there is a ``strong" SPT invariant given by a choice of $d$-dimensional SPT protected by the internal symmetry. However, many choices of such data that would produce an anomaly-free CSPT in fact produce an anomalous MSPT. We explain why these anomalies are ubiquitous for MSPTs.

    The rest of the paper is devoted to examples and applications of the formalism. In Sec.~\ref{sec:1DCluster}, we apply our results to the symmetries of simple $\Z_N$ cluster Hamiltonians to straightforwardly reproduce a theorem of Ref.~\cite{han2023modulated} restricting the SPT data for cluster states with translation symmetry. In Sec.~\ref{sec:1DDipolar}, we classify $(1+1)$D phases protected by $\Z_N$, $\U1$, and general finite Abelian dipolar symmetries; some results are new and some reproduce results from the literature. In Sec.~\ref{sec:Other1D}, we give a number of applications to (1+1)D systems with a variety of modulated symmetries, including an example with point group symmetry. In Sec.~\ref{sec:2DDipolar}, we classify (2+1)D phases protected by $\Z_N$ and $\U1$ dipolar symmetries. We summarize the results of our dipolar SPT classification in Table~\ref{tab:dipoleClassification}. Finally, we give some conclusions and open directions in Sec.~\ref{sec:conclusions}.
    
    \begin{table}[]
        \centering
        \begin{tabular}{lc||c|c}
        Symmetry& Dimension & Strong indices & Weak indices \\
        \hline
            $\Z_N^{(Q)} \times \Z_N^{(x)}$ & & &\\
                & (1+1)D & $\Z_N$ & N/A\\
                & (2+1)D & $\Z_N^3$ & N/A\\
            $\Z_N^{(Q)}\times \Z_N^{(x)} \times \Z_N^{(y)}$ &&&\\
                & (2+1)D & $\Z_N^7$ & N/A\\ 
            $\Z_N^{(Q)}\times \Z_N^{(x)} \rtimes \mathbb{T}$  &  & &\\
                & (1+1)D & $\Z_N$ & $\Z_N$\\
                & (2+1)D & $\Z_N\times \Z_{(2,N)}$ & $\Z_N^3$\\
            $\Z_N^{(Q)}\times \Z_N^{(x)} \times \Z_N^{(y)} \rtimes \mathbb{T}$ &&&\\
                &(2+1)D & $\Z_N^4$ & $\Z_N^4$\\
            $\U1^{(Q)} \times \U1^{(x)} \rtimes \mathbb{T}$ &&&\\
                & (1+1)D & 0 & $\Z$\\
                & (2+1)D & $\Z$ & $\Z$\\
            $\U1^{(Q)} \times \U1^{(x)} \times \U1^{(y)} \rtimes \mathbb{T}$ &&&\\
                & (2+1)D & $\Z^3$ & $\Z$ 
        \end{tabular}
        \caption{Classification of dipolar SPTs in (1+1)D and (2+1)D on the square lattice. $Q$ refers to charge conservation, $x$ and $y$ refer to conservation of $x$ and $y$ dipole moment respectively, $\mathbb{T}$ refers to discrete (square lattice) translation symmetry, and in (2+1)D, is always assumed to include translations in both the $x$ and $y$ directions. $(a,b)$ means the greatest common divisor of $a$ and $b$.}
        \label{tab:dipoleClassification}
    \end{table}

\section{Modulated symmetries}

\label{sec:modulated}

While the term ``modulated symmetry" has been used heavily in the literature, we will make a precise definition that matches Ref.~\cite{PaceGaugingModulated}, carefully defines the scope of this paper and that, we hope, will be useful in sharply clarifying the distinction between modulated and unmodulated symmetries.

Given a group of internal symmetries $G_{int}$ which do not modify space and a group of spatial symmetries $G_{sp}$, we define the internal symmetries to be \textit{modulated} if the total symmetry group $G$ is 
\begin{equation}
    G = G_{int} \rtimes G_{sp}.
\end{equation}
The key point is the semidirect product. Specifically, given an element $S \in G_{sp}$, there is an action
\begin{equation}
    S:G_{int} \rightarrow G_{int}
    \label{eqn:spaceGroupAction}
\end{equation}
by conjugation $g \rightarrow SgS^{-1}$ for $g \in G_{int}$, i.e., elements of $G_{sp}$ implement automorphisms of $G_{int}$. Denote this action
\begin{equation}
    S(g) = SgS^{-1}.
    \label{eqn:modulation}
\end{equation}
Note that we overload the notation; $S$ can represent either a space group element or the corresponding automorphism of $G_{int}$. The distinction should be clear from context. If the semidirect product is trivial, i.e., $S(g)=g$ for all $g \in G_{int}$ and all $S \in G_{sp}$, then we say the symmetry is unmodulated.

\begin{figure}
    \centering
    \begin{subfigure}[b]{0.7\textwidth}
        \includegraphics[width=\textwidth]{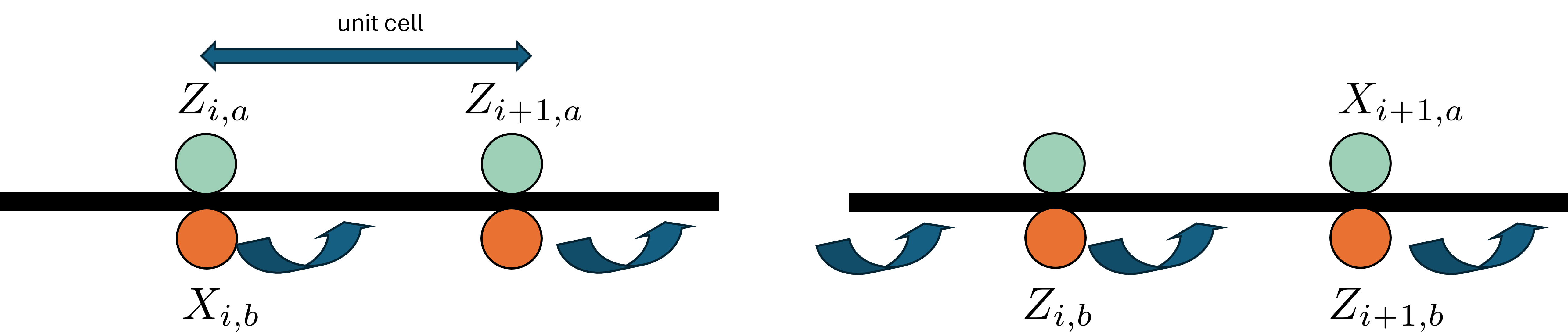}
        \caption{}
        \label{fig:UnmodulatedCluster}
    \end{subfigure}
    \begin{subfigure}[b]{0.4\textwidth}
        \includegraphics[width=\textwidth]{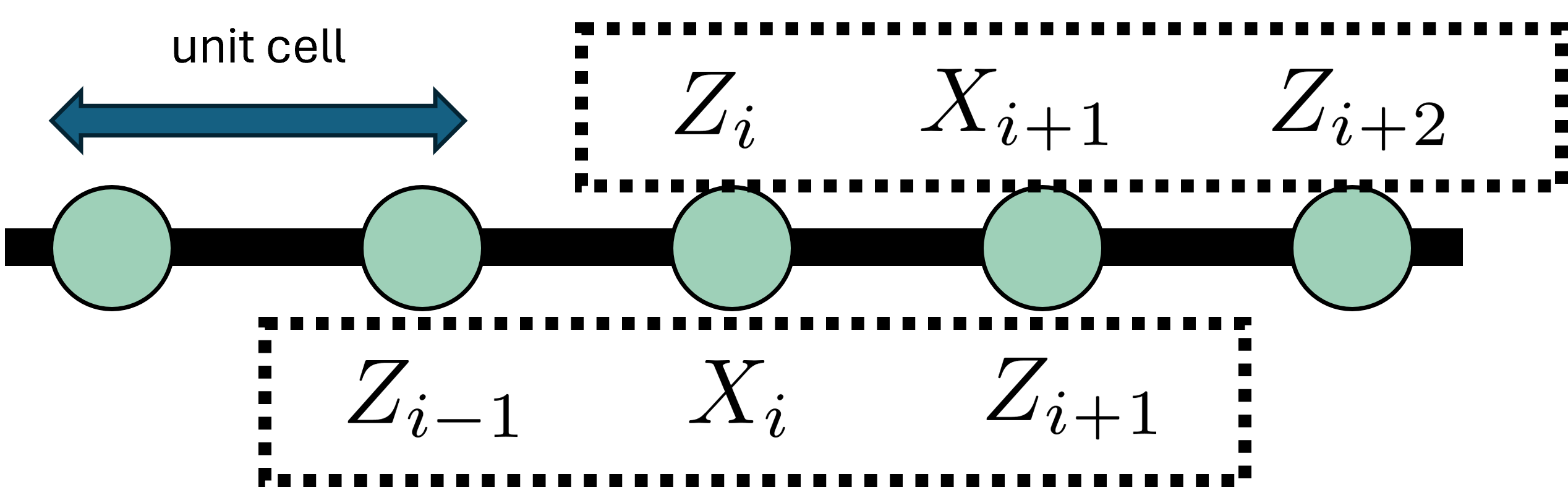}
        \caption{}
        \label{fig:ClusterH}
    \end{subfigure}
    \caption{(a) The two different coupling terms in the Hamiltonian $H_1$ (Eq.~\ref{eqn:clusterH1}), which has an unmodulated symmetry. Each site has two spin-1/2 degrees of freedom (teal and orange circles), and each Hamiltonian term couples three spins. Blue arrows indicate that by moving the ``$b$" spins (orange), the model Eq.~\ref{eqn:clusterH1} becomes exactly the model Eq.~\ref{eqn:clusterH2}, up to the translation symmetry that is imposed (and the relabeling of sites). $\Z_2$ charge is conserved on the teal ``$a$" spins alone and on the orange ``$b$" spins alone. (b) Coupling terms in the Hamiltonian $H_2$ (Eq.~\ref{eqn:clusterH2}), which has a modulated symmetry. Each site has one spin-1/2 degree of freedom (circles), and each Hamiltonian term (boxed) couples three spins. $\Z_2$ charge is conserved on the odd sites alone and on the even sites alone.}
    \label{fig:ClusterHamiltonians}
\end{figure}

To illustrate the concept, we will compare two closely related models.
Consider first a model with two spin-1/2 degrees of freedom per site; we call the corresponding Pauli operators $X_{i,a}$, $X_{i,b}$, $Z_{i,a}$, and $Z_{i,b}$, where $i$ runs over all sites and $a$ and $b$ label the two types of spins. Then consider the Hamiltonian
\begin{equation}
    H_1 = -\sum_i Z_{i,a}X_{i,b}Z_{i+1,a} - \sum_i Z_{i,b}X_{i+1,a}Z_{i+1,b} - h \sum_{i} (X_{i,a}+X_{i,b})
    \label{eqn:clusterH1}
\end{equation}
For simplicity of exposition we will work with an infinite system. The first two terms of $H_1$ are shown pictorially in Fig.~\ref{fig:UnmodulatedCluster}. We focus on three particular symmetries of this model: translation symmetry $T_1$, and two internal $\Z_2$ symmetries $Q_a,Q_b$ defined by:
\begin{align}
    Q_a &= \prod_i X_{i,a}\\
    Q_b &= \prod_i X_{i,b}
    \label{eqn:clusterSymmH1}
\end{align}
Clearly $[Q_a,Q_b]=[T,Q_a]=[T,Q_b]=0$. Hence this symmetry is unmodulated and forms the group $\Z_2 \times \Z_2 \times \Z$, where the last copy of $\Z$ represents translation.

By comparison, consider a model with one spin-1/2 degree of freedom per site with Hamiltonian
\begin{equation}
    H_2 = -\sum_i Z_{i-1}X_i Z_{i+1} - h \sum_i X_i
    \label{eqn:clusterH2}
\end{equation}
This model is the standard 1+1D $\Z_2 \times \Z_2$ cluster state Hamiltonian and is shown pictorially in Fig.~\ref{fig:ClusterH}; it has translation symmetry $T_2$ and two internal $\Z_2$ symmetries
\begin{align}
    Q_e &= \prod_{i \text{ even}} X_{i}\\
    Q_o &= \prod_{i \text{ odd}} X_{i}
\end{align}
Clearly $[Q_e,Q_o]=0$, but now
\begin{equation}
    T_2 Q_e T_2^{-1} = Q_o \quad \text{and } T_2 Q_o T_2^{-1} = Q_e
\end{equation}
The spatial symmetry group $\Z$ now has a nontrivial action on the internal symmetry group $\Z_2 \times \Z_2$ since the generator of $\Z$ swaps the copies of $\Z_2$. This symmetry is modulated and forms the group $\Z_2 \times \Z_2 \rtimes \Z$.

Clearly, one can interleave the $b$ spins in $H_1$ to obtain $H_2$, as shown in Fig.~\ref{fig:ClusterHamiltonians}, which maps
\begin{equation}
    (X_{i,a},Z_{i,a}) \rightarrow (X_{2i},Z_{2i}),\quad (X_{i,b},Z_{i,b}) \rightarrow (X_{2i+1},Z_{2i+1}).
\end{equation}
Observe that doing so actually maps $Q_a \rightarrow Q_e$ and $Q_b \rightarrow Q_o$. Hence $H_1$ and $H_2$ are identical at the level of Hilbert space (at least if we instead take appropriately sized finite systems), Hamiltonian, internal symmetry, and spatial symmetry group; the only difference is the action of translation on the internal symmetries. 

In fact, when doing this interleaving procedure, $H_1$ has an additional symmetry $T_2$ such that $T_2^2=T_1$ and $T_2$ matches its corresponding operator for $H_2$. We imagine distinguishing $H_1$ and $H_2$ by allowing perturbations to break that symmetry; we have tuned to the point with this additional symmetry in order to emphasize the equivalence of these models apart from translation symmetry. The rest of this paper will be devoted to showing that imposing the ``finer" translation symmetry $T_2$ has highly nontrivial consequences; to that end, we will now argue that the presence of such spatial symmetries is \textit{required} in order to meaningfully make a distinction between modulated and unmodulated symmetries.

\subsection{Modulated symmetries are only meaningful with spatial symmetries}

Even if internal symmetries are modulated, it is possible to break spatial symmetries without breaking the internal symmetries or modifying the generators. For example, we could choose to stagger the coefficients in front of either term in Eq.~\ref{eqn:clusterH2}; this would break the translation symmetry down to the same translation symmetry as in Eq.~\ref{eqn:clusterH1}. One could further introduce more arbitrary coefficients in front of each term in a way that completely breaks translation symmetry, but as long as we do not add additional terms to the Hamiltonian, the symmetries $Q_a$ and $Q_b$ (or $Q_e$ and $Q_o$) will be preserved. Clearly, once translation symmetry is broken, $H_1$ and $H_2$ are exactly equivalent in their Hilbert spaces (as long as we match the system size appropriately), their Hamiltonians, their symmetries, and their symmetry-allowed perturbations. There is nothing left to distinguish the two models. However, in the presence of the translation symmetries we have discussed, while the internal and spatial symmetry groups are identical\footnote{Strictly speaking, on a finite system, if we match the Hilbert space then the system size is different and the spatial symmetry groups are different. On an infinite system, the spatial symmetry group is the same, but the equivalence of the Hilbert spaces is less meaningfully defined.}, the spatial symmetry operator is different, which leads to a subtly different overall symmetry group.

As such, we argue that if there is no spatial symmetry, there is no distinction between a modulated and unmodulated symmetry. In fact, all that is required to meaningfully say that we have a $G_{int}$ global internal symmetry in $(d+1)$ dimensions is that the (global) symmetry operators $U_g$ have nontrivial support on an extensive subspace of the Hilbert space and that the $U_g$ form a faithful representation of $G_{int}$. Otherwise, the spatial structure of the symmetry operators is unimportant.

The conclusion here is that, at least at the level of topological classifications, any situation in which modulated symmetries are meaningfully distinct from unmodulated internal symmetries \textit{inevitably} requires spatial symmetries. Presumably there is a more general statement, but it may be subtle; for example, upon gauging a modulated symmetry, fractons may arise~\cite{DelfinoFractonsGaugingExponential} and it is not precisely clear what role translation symmetry plays at that stage.

In the context of SPTs, this means that we always expect to be able to turn an MSPT protected by $G_{int} \rtimes G_{sp}$ into a $G_{int}$ SPT by breaking the spatial symmetries. Roughly speaking, we expect ``strong" MSPTs (those protected ``only" by $G_{int}$, in a sense which we will discuss later) to always have some corresponding unmodulated $G_{int}$ SPT. This statement will become sharper in the context of the defect network construction, to which we turn next.

\section{Defect Networks for Unmodulated Symmetries}

\label{sec:DefectNetworks}

We begin by reviewing the defect network formalism for unmodulated symmetries in the language most suitable for generalization to modulated symmetries. For simplicity of exposition, we will also focus on bosonic SPT states within group cohomology, using the usual notation $\H^k(G,\U1)$ to mean the $k$th cohomology group of a group $G$ with coefficients in $\U1$, but the results generalize immediately to fermions and general invertible phases. The construction also works for non-invertible phases, but its use for classification in that case is less clear.

We will give a recipe for constructing a general CSPT in $d$ spatial dimensions. Specifically, we have in mind constructing a Hamiltonian whose ground state realizes a CSPT state, but the discussion applies equally well to constructing a ground state wavefunction for a CSPT state.

The main idea of the defect network construction is to associate SPTs of various dimensions protected by internal symmetries to different submanifolds in space. By consistently and symmetrically gluing that data together to form a gapped state, we construct a CSPT.

More concretely, the procedure for constructing a defect network is as follows.
\begin{enumerate}
    \item Construct a cellulation of space such that every element of $G_{sp}$ maps every cell to another cell of the same dimension. If using the defect network to classify CSPTs, add the additional constraints:
        \begin{itemize}
            \item[a)] Each $d$-cell occupies a fundamental domain of the space group.
            \item[b)] Each inequivalent dimension $<d$ Wyckoff cell of the space group is occupied by a cell of the appropriate dimension. We define $G_\Sigma \subset G_{sp}$ to be the ``little group" of the cell $\Sigma$, that is, the spatial symmetries which map $\Sigma$ to itself (possibly with an orientation reversal).
            \item[c)] No symmetries in $G_{sp}$ ``accidentally" preserve any cells, i.e., if there is a $G_{sp}$-symmetric deformation of the cellulation which shrinks some cell's $G_{\Sigma}$, perform that deformation.
        \end{itemize}
    \item To one $d-$cell, associate an $(d+1)$-dimensional $G_{int}$-SPT $[\omega] \in \H^{d+1}(G_{int},\U1)$.
    \item Apply space group symmetries to place $G_{int}$-SPTs on all of the $d$-cells.
    \item For each $G_{sp}$ orbit of $(d-1)$-cells, choose a reference cell $\Sigma_{d-1}$. Couple the two $d-$cells that border $\Sigma_{d-1}$ so that $\Sigma_{d-1}$ is occupied by a gapped state symmetric under $G_{int} \times G_{\Sigma_{d-1}}$. If such a coupling exists, there may be many choices; such a choice associates certain data $[\mu] \in \H^{d}(G_{int} \times G_{\Sigma_{d-1}},\U1)$ to each $(d-1)$-cell $\Sigma_{d-1}$.
    \item Use space group symmetries to place the above coupling on every $(d-1)$-cell.
    \item Repeat steps 4-5 on each lower-dimensional cell, all the way down to the $0$-cells. If at any point a gapped symmetric state is not possible, conclude that there is an anomaly associated with the higher-dimensional data.
\end{enumerate}

We use the terminology that $[\omega]$ is the ``strong SPT data" (since it specifies a $G_{int}$ SPT in the absence of spatial symmetries) and the data $[\mu]$ and its lower-dimensional analogues together form the ``weak SPT data". We presently elaborate on each step.

Step 1: A cellulation consists of a collection of $d$-dimensional disks (``$d$-cells"), a collection of $(d-1)$-dimensional boundaries between those disks (``$(d-1)$-cells"), a collection of $(d-2)$-dimensional boundaries between the $(d-1)$-cells, and so forth, all the way down to a collection of 0-cells, such that when glued together, they cover all of space. We will often refer to a cell of dimension $k$ with the notation $\Sigma_k$. We always demand that the cellulation itself is symmetric under $G_{sp}$.

Regarding classification, the condition (a) ensures that the defect network is ``coarse enough" that the orbit of a single $d$-cell under $G_{sp}$ covers all of space, but ``fine enough" that $G_{sp}$ acts faithfully on the collection of $d$-cells.

The condition (b) ensures that if symmetry is forced to map some $k$-dimensional structure to itself, then some $k$-cell sees that symmetry as an effective \textit{internal} symmetry. This allows us to bootstrap our SPT classification for internal symmetries to spatial symmetries. We are using the terminology for Wyckoff positions a bit loosely; we refer to a dimension $k$ ``Wyckoff cell" as a Wyckoff position which can be continuously deformed in $k$ dimensions while remaining in the same Wyckoff class. As an example, in (2+1)D with wallpaper group $pm$ (translations plus a reflection axis), each reflection axis is a dimension 1 Wyckoff cell.

The condition (c) ensures that there is no data associated to the defect network which is protected by an ``accidental" symmetry, in the sense that all data should be robust under all $G_{sp}$-preserving deformations.

In general, classification requires classifying inequivalent defect networks, where ``equivalent" networks are related to each other by a process of symmetrically creating defects which fuse to the vacuum, then symmetrically deforming those defects. For the spatial symmetries we discuss in this paper, these equivalences do not appear in the unmodulated case; for such examples, see Refs.~\cite{ElseThorngrenDefectNetworks,HuangBlockStates}. However, we will see that many equivalences will appear in the modulated case; we discuss these in Sec.~\ref{sec:equivalences}.

Step 2 does not require any additional comment.

Step 3: Since all elements of $G_{sp}$ commute with $G_{int}$, this step places the \textit{same} SPT $[\omega]$ on each cell. We can then temporarily choose (generically gapless) $G_{int}$-symmetric boundary conditions for each $d$-cell. Then, since the $d$-cells are not yet coupled, the whole system has internal symmetry $G_{int}^{N_d}$, where $N_d$ is the number of $d$-cells because each $d$-cell is an independent system with a $G_{int}$ symmetry. For shorthand, we will call one of these copies of $G_{int}$ a ``local copy" since it only acts on a single cell. This overall symmetry is, of course, much larger than the desired symmetry of our SPT, which only has one copy of $G_{int}$.

We emphasize for later use that the SPT label $[\omega]$ determines the properties of the state as specified by the \textit{local} symmetry operators $U_g(i)$ which form the \textit{local} ($i$th) copy of $G_{int}$.

\begin{figure}
    \centering
    \begin{subfigure}[b]{0.5\textwidth}
        \includegraphics[width=\textwidth]{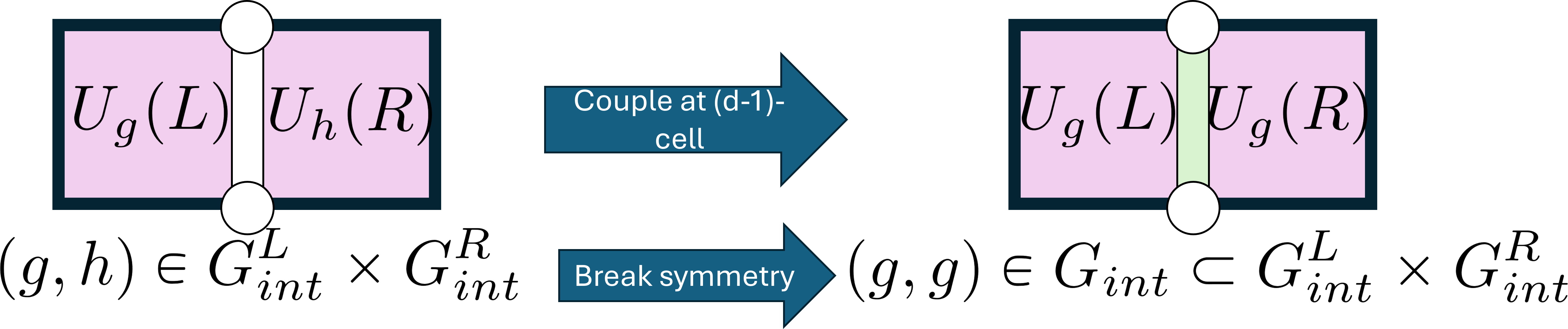}
        \caption{}
        \label{fig:UnmodulatedDMinus1Cell}
    \end{subfigure}
    \begin{subfigure}[b]{0.64\textwidth}
        \hspace{-1cm}\includegraphics[width=\textwidth]{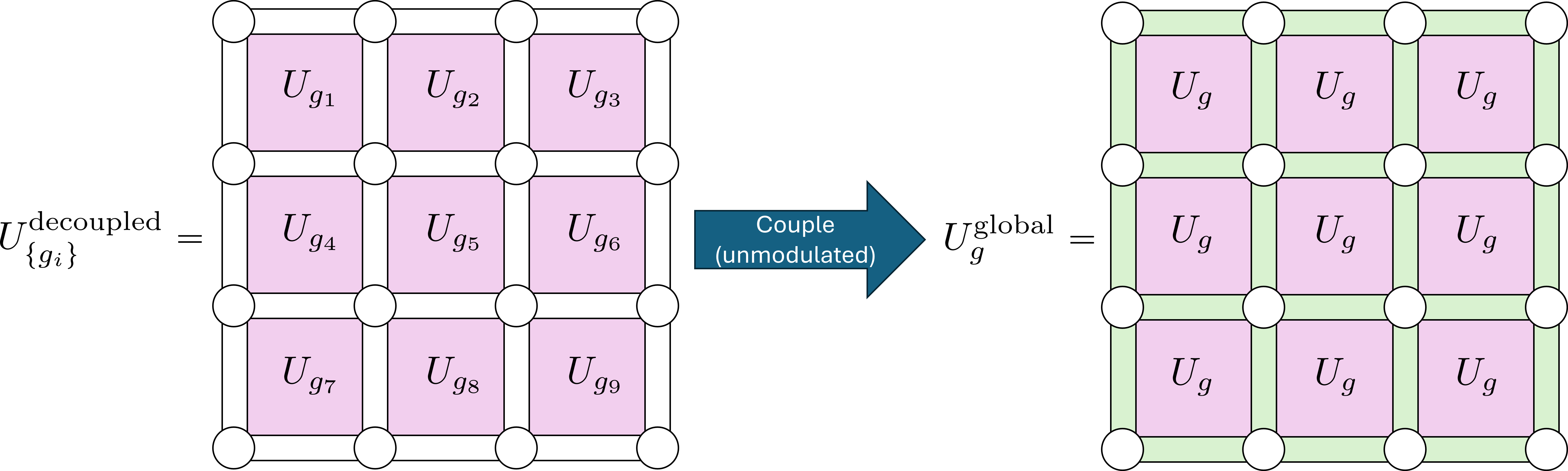}
        \caption{}
    \end{subfigure}
    \begin{subfigure}[b]{0.5\textwidth}
        \includegraphics[width=\textwidth]{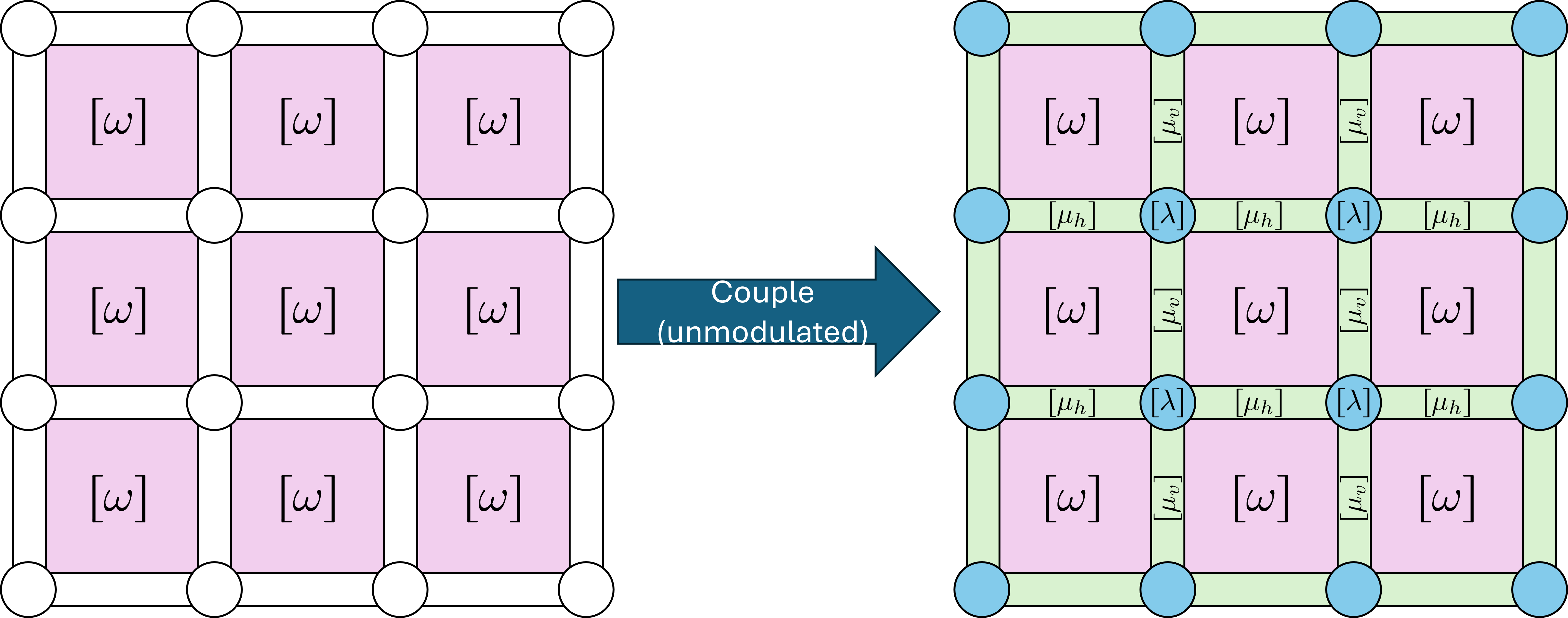}
        \caption{}
        \label{fig:UnmodulatedSquareLattice}
    \end{subfigure}
    \caption{Defect network construction for an unmodulated symmetry, shown with $d=2$ and space group consisting only of translations. (a) Two $d$-cells (pink) are coupled on the $(d-1)$-cell; the symmetry breaks from independent copies of $G_{int}$ on each $d$-cell to the diagonal subgroup. (b) Resulting global symmetry operators after performing the coupling in (a) on all $(d-1)$-cells. $N_d$ independent copies of $G_{int}$, one for each $d$-cell, breaks down to the diagonal subgroup. (c) Data assigned before and after coupling the $d$-cells on all lower-dimensional cells. The data $[\omega]$ placed on each $d$-cell is unchanged after the coupling since the global symmetry operator associated to $g \in G_{int}$ transforms each $d$-cell by its local $g$ symmetry operator $U_g$ (see (b)). The data associated to lower-dimensional cells can be different for different orbits under the space group; here, the horizontal 1-cells and vertical 1-cells form different orbits under the space group, so there is a separate choice of data $[\mu_h]$ and $[\mu_v]$  for each. The single orbit of 0-cells gets one piece of data $[\lambda]$.}
\end{figure}

Step 4: This step is shown graphically in Fig.~\ref{fig:UnmodulatedDMinus1Cell}. The $(d-1)$-cell $\Sigma_{d-1}$ has two neighboring $d$-cells (which, based on Fig.~\ref{fig:UnmodulatedDMinus1Cell}, we call the ``left" (L) and ``right" (R) $d$-cells), each of which has its own local copy of $G_{int}$. We choose the coupling to break the internal symmetry that acts on the $d$-cells down to the diagonal subgroup
\begin{equation}
    G_{int}^L \times G_{int}^L \rightarrow \lbrace U_g(L)U_g(R)|g \in G_{int}\rbrace \simeq G_{int} \subset G_{int}^L \times G_{int}^R
\end{equation}
and restore the energy gap. Here the notation $U_g(L)U_g(R)$ means we act with the same local element of $G_{int}$ on both $d$-cells that neighbor $\Sigma_{d-1}$, and we are using the $L$ and $R$ labels to distinguish local copies of $G_{int}$. Note that $G_{int} \subset G_{int} \times G_{\Sigma_{d-1}}$, so if $G_{\Sigma_{d-1}}$ is nontrivial, we can turn this into a two-step process where we first break the $G_{int}^L\times G_{int}^R$ symmetry down to the diagonal subgroup, then extend the internal symmetry of the $(d-1)$-cell to $G_{int} \times G_{\Sigma_{d-1}}$.

Such a gapped interface between the two $d$-cells involves a choice of data $[\mu]$~\cite{wang2018symmetric}, which is a trivialization of the difference of the two $(d+1)$-cocycles living on each side of the $(d-1)$-cell. This data $[\mu]$ is thus, \text{a priori}, a $d$-cochain, not a $d$-cocycle, and inequivalent data correspond to layering $(d-1)$-dimensional SPTs on top of some reference choice of data. However, by translation symmetry, the cocycles on each side of the $(d-1)$-cell are identical on the nose; therefore, $[\mu]$ is a trivialization of the identity $(d+1)$-cocycle, which is just given by a $d$-cocycle. Hence $[\mu]$ is indeed a $d$-cocycle, and distinct choices are classified by $\H^d(G_{int} \times G_{\Sigma_{d-1}},\U1)$.

Step 5; This reduces the large symmetry group $G_{int}^{N_d}$ down to its diagonal subgroup, that is, one copy of $G_{int}$.

Step 6: Given valid data (and barring any anomalies in lower dimensions), it is always possible to stack a $G_{int}\times G_{\Sigma}$ SPT on top of any given data for a $k$-cell. This means that in general, the data on a $k$-cell will form a $\H^{k+1}(G_{int}\times G_{\Sigma},\U1)$ torsor, as stacking different SPTs typically leads to distinct data. However, in general, anomalies can occur; see Ref.~\cite{ElseThorngrenDefectNetworks} for an example of a fermionic phase with an anomaly on the 0-cells. 

\section{Defect Networks for Modulated Symmetries}

\label{sec:modulatedDefect}

\subsection{Modifying the construction to introduce modulation}

We will now explain how to modify the defect network construction to obtain modulated symmetries. Steps 1-3 in Sec.~\ref{sec:DefectNetworks} are unchanged; we still cellulate space, pick a $(d+1)-$dimensional $G_{int}-$SPT, and place it on all of the $d$-cells. Importantly, we will define the action of $G_{sp}$ to be purely spatial, in the sense that it may permute the degrees of freedom on different cells, but it does not perform any internal operations on the cells. This is the same as for an unmodulated symmetry; we will introduce the modulation momentarily. We still have placed the same $G_{int}$-SPT $[\omega]$ on each $d$-cell, but crucially, the label $[\omega]$ refers to each $d$-cell's local copy of $G_{int}$.

Set the notation that under its local copy of $G_{int}$, the $d$-cell $\Sigma_d^{(i)}$ transforms under the operators $U_g(i)$ for $g \in G_{int}$.

As a technical point, for consistency with the group action in Eq.~\ref{eqn:modulation} we will take the convention that spatial symmetries act \textit{actively}, that is, they move the degrees of freedom and not the coordinates. This means that if $S\in G_{sp}$, then the action on \textit{operators} is
\begin{equation}
    S(U_g(x)) = U_g(S(x)).
\end{equation}

We introduce the modulation at the stage of coupling the $d$-cells, i.e., step 4 in Sec.~\ref{sec:DefectNetworks}, and will discuss the lower-dimensional cells later. We still need to break the $G_{int}^{N_d}$ symmetry down to a single copy of $G_{int}$, but the key observation is that spatial symmetries also \text{permute} the local copies of $G_{int}$. As such, we can introduce the modulation by breaking $G_{int}^{N_d}$ not to the diagonal subgroup, but to a subgroup which is \textit{not} invariant under permutation.

\begin{figure}
    \centering
    \begin{subfigure}[b]{0.5\textwidth}
        \includegraphics[width=\textwidth]{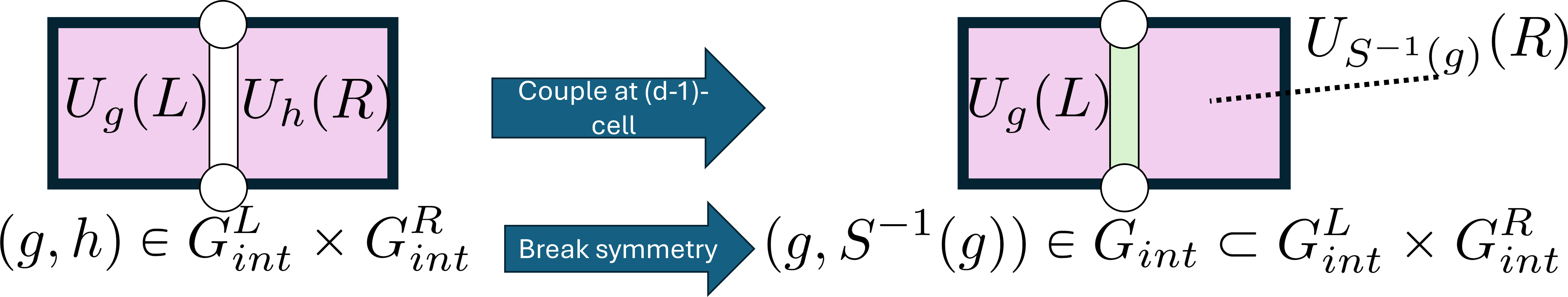}
        \caption{}
        \label{fig:ModulatedDMinus1Cell}
    \end{subfigure}
    \begin{subfigure}[b]{0.64\textwidth}
        \hspace{-1cm}\includegraphics[width=\textwidth]{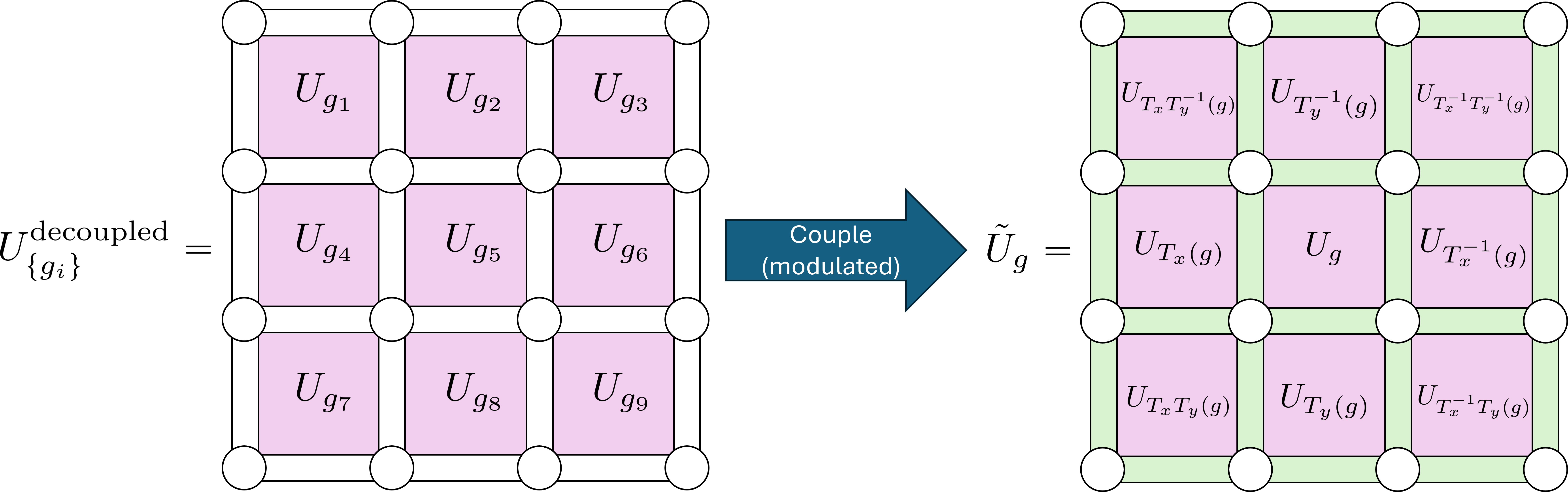}
        \caption{}
        \label{fig:ModulatedSymm}
    \end{subfigure}
    \begin{subfigure}[b]{0.5\textwidth}
        \includegraphics[width=\textwidth]{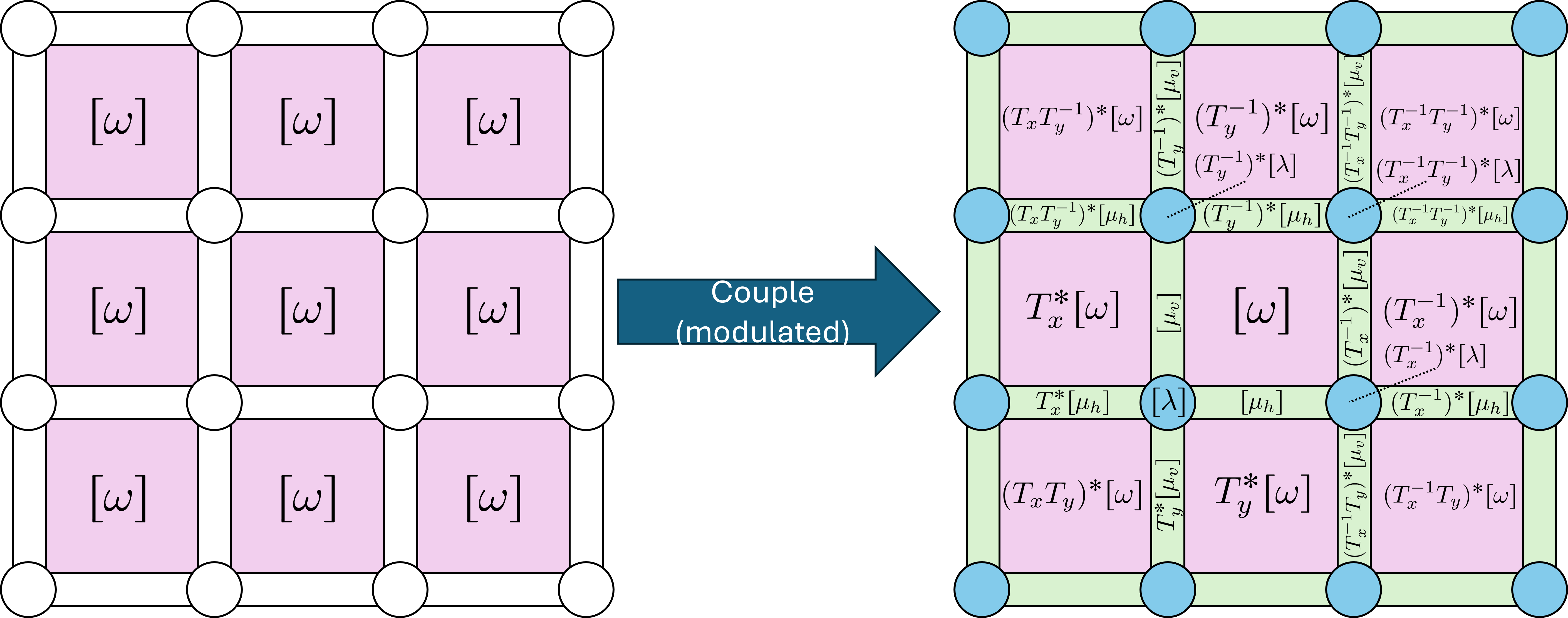}
        \caption{}
        \label{fig:ModulatedData}
    \end{subfigure}
    \caption{Defect network construction for a \textit{modulated} symmetry, shown with $d=2$ and space group consisting only of translations. (a) Two $d$-cells (pink) are coupled on the $(d-1)$-cell; the symmetry breaks from independent copies of $G_{int}$ on each $d$-cell to a \textit{non-diagonal} subgroup. Here $S$ is a space group element that maps the left $d$-cell to the right $d$-cell. (b) Resulting global symmetry operators after performing the coupling in (a) on all $(d-1)$-cells. $N_d$ independent copies of $G_{int}$, one for each $d$-cell, breaks down to a single non-diagonal $G_{int}$ subgroup. (c) Data assigned before and after coupling the $d$-cells on all lower-dimensional cells. The data $[\omega]$ placed on each $d$-cell is modified by the space group action after the coupling since the global symmetry operator associated to $g \in G_{int}$ transforms each $d$-cell by a different local symmetry operator (see (b)).}
\end{figure}

Consider a particular $(d-1)$-cell $\Sigma_{d-1}$ as shown in Fig.~\ref{fig:ModulatedDMinus1Cell}. Since each $d$-cell is a fundamental domain of the symmetry, there is a unique space group element $S$ which maps the ``left" $d$-cell $\Sigma_d^{L}$ to the ``right" $d$-cell. Then we choose the coupling on the $(d-1)$-cells to break the symmetry 
\begin{equation}
    G_{int}^L \times G_{int}^R \rightarrow \lbrace U_g(L)U_{S^{-1}(g)}(R) | g \in G_{int}\rbrace \simeq G_{int} \subset G_{int}^L \times G_{int}^R.
    \label{eqn:modulatedCoupling}
\end{equation}
Shifting the operators by $S$ changes the operator on the right-hand $d$-cell from $U_{S^{-1}(g)}$ to $U_{g}$, that is, it acts by $S$ on the $g$ label, which is what we want for the modulation. This is the reason that $S$ appears with an inverse. This is a conceptually similar construction to that used in Ref.~\cite{JiangLuGeneralizedLSM}, which focuses on systems with onsite projective representations of the symmetries.

Now, using space group symmetries, place the same coupling on all $(d-1)$-cells in the orbit of $\Sigma_{d-1}$.

We now make three crucial claims about the properties and consequences of this coupling. We make these claims in the logical order required to prove them, but the second claim is the most important.

\begin{enumerate}
    \item Choosing couplings Eq.~\ref{eqn:modulatedCoupling} on every $(d-1)$-cell correctly breaks the global symmetry action on the $d$-cells down to $G_{int} \rtimes G_{sp}$.
    \item Such a coupling is only possible while maintaining the gap and $G_{int}$ symmetry when the condition
\begin{equation}
    S^\ast[\omega]=[\omega]
    \label{eqn:modulatedAnomaly}
\end{equation}
is satisfied, where $S^\ast$ is the pullback of the group action Eq.~\ref{eqn:spaceGroupAction} to cohomology
\begin{equation}
    S^\ast : \H^{d+1}(G_{int},\U1)\rightarrow \H^{d+1}(G_{int},\U1),
\end{equation}
defined by (using inhomogeneous cocycles)
\begin{equation}
    \left(S^\ast\omega\right)(g_0,g_1,\ldots,g_d) = \omega\left(S(g_0),S(g_1),\ldots, S(g_d)\right).
\end{equation}
This is one of the main results of this paper; when Eq.~\ref{eqn:modulatedAnomaly} is not satisfied, there is a 't Hooft anomaly associated with the data $[\omega]$ for the modulated symmetry.
\item Distinct couplings on a $(d-1)$-cell $\Sigma_{d-1}$ are given by different classes $[\mu] \in C^d(G_{\Sigma_{d-1}},\U1)/B^d(G_{\Sigma_{d-1}})$ which trivialize the coboundary $S^\ast\omega/\omega$.
\end{enumerate}
Here $C^d(G_{\Sigma_{d-1}},\U1)$ is the set of $d$-cochains with $\U1$ coefficients and $B^d(G_{\Sigma_{d-1}},\U1)$ is the set of $d$-coboundaries with $\U1$ coefficients. Let us take these claims one by one.

\subsubsection{Modulated symmetry}

What global symmetry remains after performing this coupling? If we act with a local symmetry $U_g$ on some reference $d$-cell, then we can determine the action on every neighboring $d$-cell by acting on $g$ with the appropriate $S \in G_{sp}$ every time we pass through a $(d-1)$-cell. An example for $d=2$ with only translation symmetry is shown in Fig.~\ref{fig:ModulatedSymm}. To be more precise, let $\Sigma_d^{(i)}$ label the different $d$-cells, where $i$ ranges from $0$ to $N_d-1$, and suppose that $S_i$ is the (unique, since the $d$-cells are all fundamental domains) element of $G_{sp}$ which maps $\Sigma_d^{(0)}$ to $\Sigma_d^{(i)}$. Then these couplings break the symmetry
\begin{equation}
    G_{int}^{N_d} \rightarrow G_m = \left\lbrace \tilde{U}_g = \bigotimes_{i=0}^{N_d-1}U_{S^{-1}_i(g)}(\Sigma_d^{(i)}) \bigg| g \in G_{int}\right\rbrace \simeq G_{int},
    \label{eqn:GmBreaking}
\end{equation}
where the $m$ subscript stands for ``modulated."

We also need to check that $G_{sp}$ has the correct action on $G_m$, that is, the full symmetry group is $G_m \rtimes G_{sp}$. 

Let $T \in G_{sp}$. $T$ permutes the $d$-cells; we denote $T\left(\Sigma_d^{(i)}\right)=\Sigma_d^{T(i)}$. This permutation is what causes $T$ to act on $G_m$. For any element $g \in G_m$,
then, $T$ acts by
\begin{align}
    T(\tilde{U}_g) &= T\left(\bigotimes_{i=0}^{N_d-1}U_{S^{-1}_i(g)}(\Sigma_d^{(i)})\right)\\
    &= \bigotimes_{i=0}^{N_d-1}U_{S^{-1}_i(g)}(T(\Sigma_d^{(i)}))\\
    &= \bigotimes_{i=0}^{N_d-1}U_{S^{-1}_i(g)}(\Sigma_d^{T(i)})\\
    &= \bigotimes_{i=0}^{N_d-1}U_{S^{-1}_{T^{-1}(i)}(g)}(\Sigma_d^{(i)})
\end{align}
where in the last line we reindexed $i\rightarrow T^{-1}(i)$.
By definition, $S^{-1}_{T^{-1}(i)} \in G_{sp}$ maps $\Sigma_d^{T^{-1}(i)}$ to $\Sigma_d^{(0)}$. But so does the composition $S^{-1}_iT \in G_{sp}$. Since each $d$-cell is a fundamental domain of $G_{sp}$, this means that these two group elements are equal, that is,
\begin{equation}
    S^{-1}_{T^{-1}(i)}=S^{-1}_iT.
\end{equation}
Therefore,
\begin{align}
    T(\tilde{U}_g) &= \bigotimes_{i=0}^{N_d-1}U_{S^{-1}_iT(g)}(\Sigma_d^{(i)}) \\
           &= \tilde{U}_{T(g)}
\end{align}
which is the desired action of $T$.

\subsubsection{Anomaly-free condition}

Consider a symmetry operator $\tilde{U}_g \in G_m$, for example the one shown in Fig.~\ref{fig:ModulatedSymm}. Observe that $\tilde{U}_g$ does \textit{not} transform every $d$-cell by $g$ under the \textit{local} copy of $G_{int}$; instead, $\tilde{U}_g$ transforms $\Sigma_d^{(i)}$ by $S_i^{-1}(g)$. But recall that the cocycle $\omega(g_0,\ldots,g_d)$ tells us about the SPT placed on each $d$-cell when the $g_i$ are valued in that cell's \textit{local} copy of $G_{int}$. That means that if we consider the SPT data associated to each $d$-cell where the group labels $g_0,\ldots,g_d$ are valued in the \textit{modulated} symmetry, $\Sigma_d^{(i)}$ is associated with the data 
\begin{equation}
    \omega\left(S_i^{-1}(g_0),\ldots,S_i^{-1}(g_d)\right) = (S_i^{-1})^\ast \omega(g_0,\ldots,g_d)
\end{equation}
Therefore, comparing the data assigned to two neighboring $d$-cells under the \textit{modulated} symmetry, as in Fig.~\ref{fig:ModulatedData}, we see that they only carry the same SPT data under the \textit{modulated} symmetry (and therefore allow a gapped, symmetric boundary between them) if
\begin{equation}
    S_i^\ast[\omega] = [\omega],
\end{equation}
for every $i$, which is equivalent to Eq.~\ref{eqn:modulatedAnomaly}.

The argument as we have stated it is actually implicitly reliant on assuming $G_{\Sigma_{d-1}}$ is trivial, that is, there is no additional effective internal symmetry on the $(d-1)$-cells. We give a fully general (but more technical) argument in Sec.~\ref{subsec:extendedSymm}.

\subsubsection{Data on the \texorpdfstring{$(d-1)$}{(d-1)}-cell}

Based on our discussion above, the data we need to specify on each $(d-1)$-cell $\Sigma_{d-1}$ is a gapped interface between the SPTs given by the cocycles $\omega$ and $S^\ast\omega$. If $G_{\Sigma_{d-1}}$ is trivial, then the solution to this problem is well-understood~\cite{wang2018symmetric}; all we need is a cochain $\mu \in C^d(G_{int},\U1)$ such that
\begin{equation}
    \frac{S^\ast \omega}{\omega} = d\mu
    \label{eqn:topDimTrivialization}
\end{equation}
which must exist if there is no anomaly, i.e., if Eq.~\ref{eqn:modulatedAnomaly} is satisfied. The difference between two cochains $\mu$ and $\mu'$ which both satisfy Eq.~\ref{eqn:topDimTrivialization} is nontrivial in $\H^d(G_{int},\U1)$ if and only if $\mu$ and $\mu'$ represent topologically distinct gapped boundaries. Hence the data associated to the $(d-1)$-cell is a class $[\mu] \in C^d(G_{int},\U1)/B^d(G_{int},\U1)$ which satisfies Eq.~\ref{eqn:topDimTrivialization}. This means that the data on the $(d-1)$-cell forms an $\H^d(G_{int},\U1)$ torsor.

Unlike the unmodulated case, there is no guarantee in general that $S^\ast\omega=\omega$ as cocycles, only that they are equal as cohomology classes. As such, we cannot always choose a canonical representative of the class $[\omega]$ such that the left-hand-side of Eq.~\ref{eqn:topDimTrivialization} is the trivial cocycle 1. We are generically stuck with the torsorial structure, rather than saying that $\H^d(G_{int},\U1)$ classifies the $(d-1)$-cell data. We will occasionally elide this distinction in our examples. 

We argue in Sec.~\ref{subsec:extendedSymm} that for a general $G_{\Sigma_{d-1}}$, the classification of $(d-1)$-cell data forms a $\H^d(G_{int}\rtimes G_{\Sigma_{d-1}},\U1)$ torsor.

\subsection{Interpretation of the anomaly}
The anomaly-free condition Eq.~\ref{eqn:modulatedAnomaly} is one of the key results of this paper. We will use it extensively in the examples in subsequent sections; we pause to give some interpretation.

As with a usual 't Hooft anomaly, for a given set of input data, the anomaly Eq.~\ref{eqn:modulatedAnomaly} is an obstruction to constructing a gapped phase which is symmetric under the full symmetry group, the internal symmetry acts onsite, and the onsite generators obey a \textit{linear} representation of the symmetry algebra.

Notice that the non-anomalous strong SPT data for the modulated symmetry is a \textit{subset} of the strong SPT data for the corresponding unmodulated symmetry (which is just $G_{int}$ SPT data). We also see that while the basic recipe for modulated and unmodulated CSPTs is the same, anomalies arise much more dramatically and frequently for modulated CSPTs than for unmodulated ones, even if the comparison keeps the internal and spatial symmetry groups fixed. Anomalies for unmodulated CSPTs more often appear in higher dimensions and require larger symmetry groups than pure translations, but in the modulated case, as we will see, just translation symmetry is often sufficient to lead to anomalies. 

\begin{figure}
    \centering
    \includegraphics[width=0.4\linewidth]{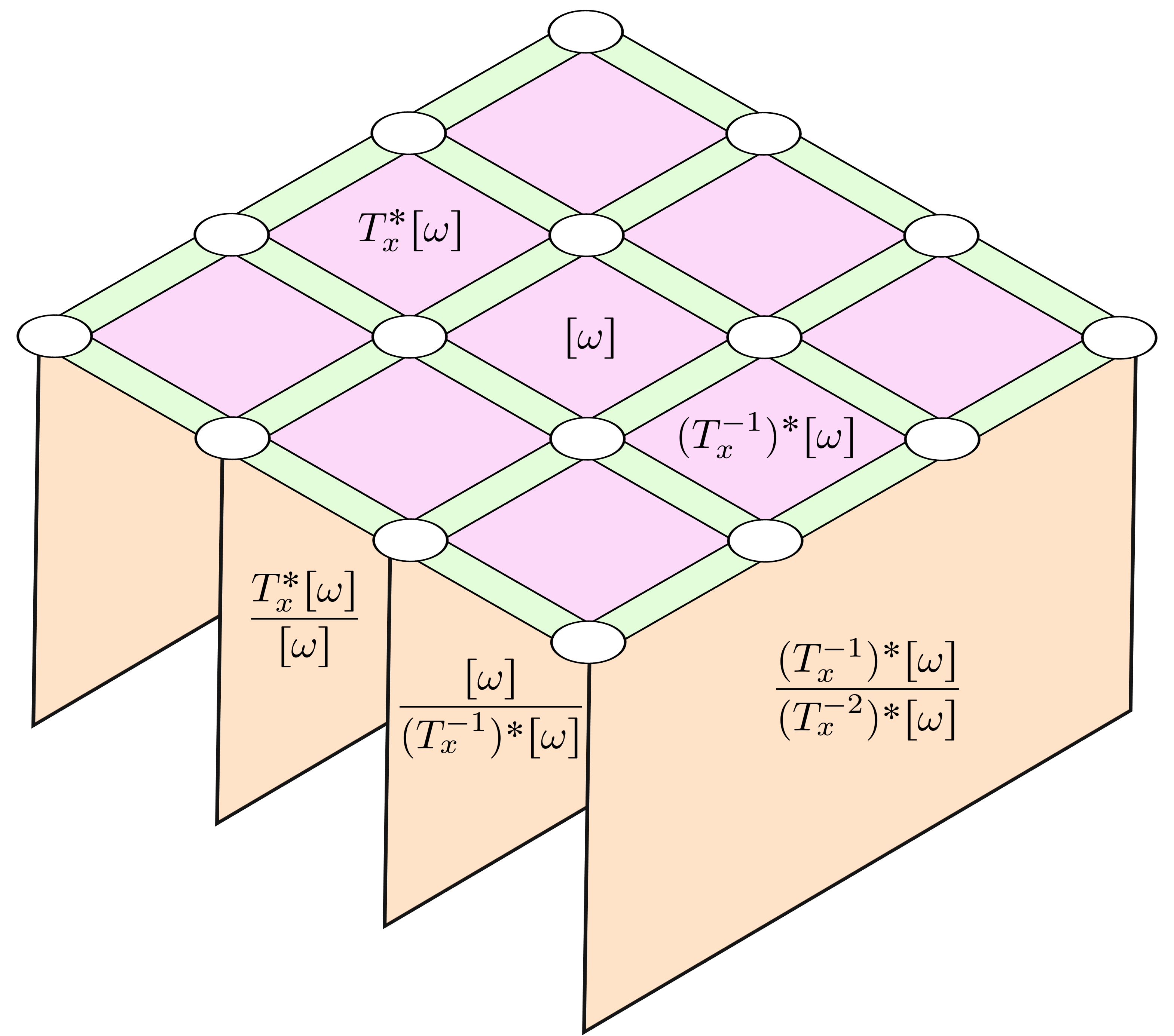}
    \caption{Anomaly cancellation in a (2+1)D modulated SPT (pink and green defect network) by modulated weak SPTs in (3+1)D. For simplicity of presentation we assume that $T_y^\ast$ acts trivially on $[\omega]$. If Eq.~\ref{eqn:modulatedAnomaly} is not satisfied for $S=T_x$, then the failure $T_x^\ast[\omega]/[\omega]$ of the anomaly condition specifies a (2+1)D $G_{int}$ SPT (orange) whose edge is placed on a reference 1-cell. The other 1-cells are decorated with edges of SPTs (also orange) according to the space group action.}
    \label{fig:anomalyCancellation}
\end{figure}

This anomaly can be canceled via a weak MSPT living in $(d+1)$ spatial dimensions. Specifically, the anomaly is given by a $(d+1)$-cocycle $ S^\ast\left([\omega]\right)/[\omega]$ (using multiplicative notation for the cohomology classes), and thus corresponds to the edge mode of a $d$-dimensional $G_{int}$ SPT. If the edge of such an SPT is placed on the reference $(d-1)$-cell and then extended to other cells by space group operations, then the anomaly is canceled, see Fig.~\ref{fig:anomalyCancellation}. Note that the anomaly is in general modulated, that is, translating the anomaly requires the action of the space group, which can change its cohomology class.

It is entirely possible that the onsite generators of the symmetry obey a projective representation of the symmetry group. If so, in (1+1)D, the anomaly condition Eq.~\ref{eqn:modulatedAnomaly} can be reinterpreted as a constraint on the projective representations compatible with the input strong SPT data, which has the flavor of a Lieb-Schultz-Mattis-Oshikawa-Hastings-type constraint~\cite{LSM,OshikawaLSM,HastingsLSM,ChengLSM}. In higher dimensions, this constraint looks more exotic. This interpretation will be explored in upcoming work~\cite{bulmashPace}.

\subsection{Importance of spatial symmetries}

Clearly the spatial symmetries play a key role in the construction, but let us see explicitly how the anomalies disappear if we break spatial symmetries. The key point is that in the absence of spatial symmetries, we are not required to place the same initial SPT $[\omega]$ (under the local copy of $G_{int}$) on every $d$-cell. In that case, we can place, for example, $S_i^\ast[\omega]$ on the $d$-cell $\Sigma_d^{(i)}$. If we do so, then once we couple the $d$-cells, under the modulated symmetry $G_m$ every $d$-cell $\Sigma_d^{(i)}$ contains the data 
\begin{equation}
    (S_i^{-1})^\ast S_i^\ast [\omega] = [\omega],
\end{equation}
and we are guaranteed the existence of a gapped, symmetric boundary. The spatial symmetry, then, forces us into the nontrivial anomaly-free condition Eq.~\ref{eqn:modulatedAnomaly}; breaking the spatial symmetry allows us to modify the construction and removes the anomaly condition on $[\omega]$, even if we do not modify the structure of the global internal symmetry operators.

\subsection{Lower-dimensional cells}
\label{sec:LowerCell}

We also need to place data on the codimension $>1$ cells so that the system is gapped everywhere, not just on the $d$- and $(d-1)$-cells. This is done in analogy to the $(d-1)$-cells, but the details of the procedure will vary somewhat depending on the cell structure. For later use, we will give the example where the 2-cells form a 2D square lattice and there is only translation symmetry; the results generalize. The cellulation and data we use are illustrated in Fig.~\ref{fig:ModulatedData}.

Suppose the state on the 2-cells is given by the cocycle $\omega \in Z^3(G_{int},\U1)$. Then a gapped interface between two neighboring 2-cells is specified by a 2-cochain $\mu$ as follows:
\begin{equation}
    d\mu = \frac{T^\ast \omega}{\omega},
\end{equation}
which is guaranteed to be possible provided $\omega$ satisfies Eq.~\ref{eqn:modulatedAnomaly}.

We now tile this choice of $\mu$ onto all other 1-cells in the same orbit, as in Fig.~\ref{fig:ModulatedData}. For the square lattice with only translation symmetry, there are two disjoint orbits, consisting of the horizontal and the vertical 1-cells. We thus have two choices $\mu_h$ and $\mu_v$, where
\begin{align}
    d\mu_h &= T_y^\ast \omega\overline{\omega} \nonumber \\
    d\mu_v &= T_x^\ast \omega\overline{\omega}.
    \label{eqn:muTrivialization}
\end{align}
Under the global symmetry $G_m$, we should view two neighboring horizontal 1-cells as containing states specified by $\mu_h$ and $T_x^\ast \mu_h$, respectively, with a similar condition for the vertical 1-cells; in particular, if $\mu_h$ obeys Eq.~\ref{eqn:muTrivialization}, then
\begin{equation}
    d\left(T_x^\ast \mu_h\right) = T_x^\ast d\mu_h = T_x^\ast \left(T_y^\ast \omega\overline{\omega}\right) = T_x^\ast T_y^\ast \omega\overline{T_x^\ast\omega},
\end{equation}
so $T_x^\ast \mu_h$ indeed trivializes the correct cocycle on the neighboring $1$-cell.

In order to have a gapped interface at the 0-cell, where four 1-cells come together, we must then have
\begin{equation}
    \mu_h \overline{T_x^\ast \mu_h} \overline{\mu_v} T_y^\ast \mu_v =d\lambda
    \label{eqn:squareLattice0CellAnomaly},
\end{equation}
where we have chosen conventions for the orientations of the cells consistent with translation symmetry, and $\lambda$ is some 1-cochain. In particular, the left-hand side must be trivial in cohomology. One can make a consistency check that the cochain appearing on the left-hand side of Eq.~\ref{eqn:squareLattice0CellAnomaly} is indeed a cocycle by making use of Eq.~\ref{eqn:muTrivialization}.

\subsection{Equivalences of weak SPT data}
\label{sec:equivalences}

We discussed in Sec.~\ref{sec:DefectNetworks} that two naively equivalent defect networks may be related by a process of symmetrically nucleating defects from vacuum, then symmetrically deforming those defects. While this did not play a strong role in the unmodulated case, we will discuss now how they are common in the modulated case.

Consider first the case of 1+1D with $G_{sp}$ consisting only of translation symmetry. Then the data on the 0-cells forms a $\H^1(G_{int},\U1)$ torsor. We claim that if we place the data $[\phi]$ onto a reference 0-cell (which we call site 0), and place the corresponding data $(T^\ast)^n[\phi]$ on site $n$, then we should identify 
\begin{equation}
    T^\ast[\phi] \sim [\phi].
    \label{eqn:0CellEquivalence}
\end{equation}
This equivalence relation is always trivial for unmodulated symmetries, consistent with our discussion in Sec.~\ref{sec:DefectNetworks}. Mathematically, this equivalence relation turns the weak SPT data into a torsor by
\begin{equation}
    \H^1(G_{int},\U1)/(T^\ast-1)\H^1(G_{int},\U1),
\end{equation}
which is called the group of coinvariants of $\H^1(G_{int},\U1)$ under the action of $T^\ast$.

\begin{figure}
    \centering
    \includegraphics[width=0.5\linewidth]{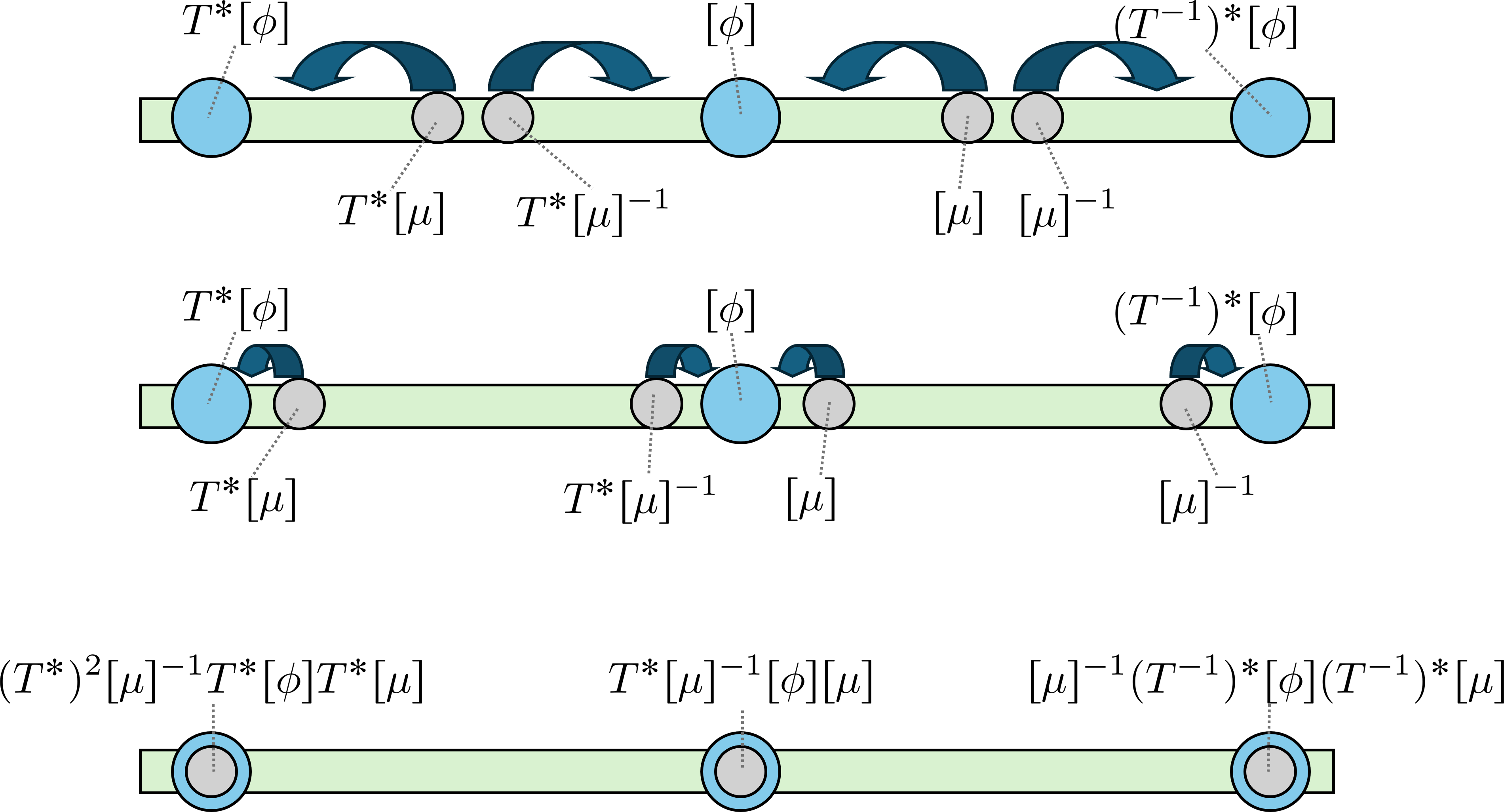}
    \caption{Physical process that produces the equivalence Eq.~\ref{eqn:0CellEquivalence}. A dipole of $G_{int}$ charge (grey circles) charge is created at the center of each 1-cell (green), and then the opposite charges are moved (blue arrows) towards opposite 0-cells (blue).}
    \label{fig:0CellEquivalence}
\end{figure}

To understand physically how this equivalence arises, consider the process in Fig.~\ref{fig:0CellEquivalence}, where we start from vacuum and pair-produce $[\mu] \in \H^1(G_{int},\U1)$ and $[\mu]^{-1}$ in the center of the 0th 1-cell. This is possible via a local, $G_{int}$-symmetric operator since these two elements fuse to the trivial element. Since elements of $\H^1(G_{int},\U1)$ are just $G_{int}$ charges, each charge is created physically by applying an operator $\mathcal{O}^{(\pm)}_0$ which is charged under $G_{int}$, where $\pm$ refers to the operator creating $[\mu]^{\pm 1}$. To do this in a translation-invariant way, we need to act by $T^n(\mathcal{O}_0)$ on 1-cell $n$; since $G_{int}$ does not commute with $T$, it follows that the $G_{int}$ charges placed on site $n$ will be $(T^\ast)^{-n}[\mu]$. However, these charges under $G_m$ are allowed to be moved adiabatically, so we can (on all the 1-cells simultaneously) move the charge $(T^n)^\ast[\mu]^{-1}$ to the right and $(T^n)^{\ast}[\mu]$ to the left. This symmetrically and adiabatically merges the combination $[\mu]^{-1}T^{\ast}[\mu]$ onto a reference 0-cell, i.e., that combination is trivial, which produces the equivalence Eq.~\ref{eqn:0CellEquivalence}.

In Appendix~\ref{app:ExactlySolvableClusterDefectNetwork}, we construct an exactly solvable defect network for the $(1+1)$D cluster state MSPT and demonstrate the above argument in a concrete model.

\begin{figure}
    \centering
    \includegraphics[width=0.9\linewidth]{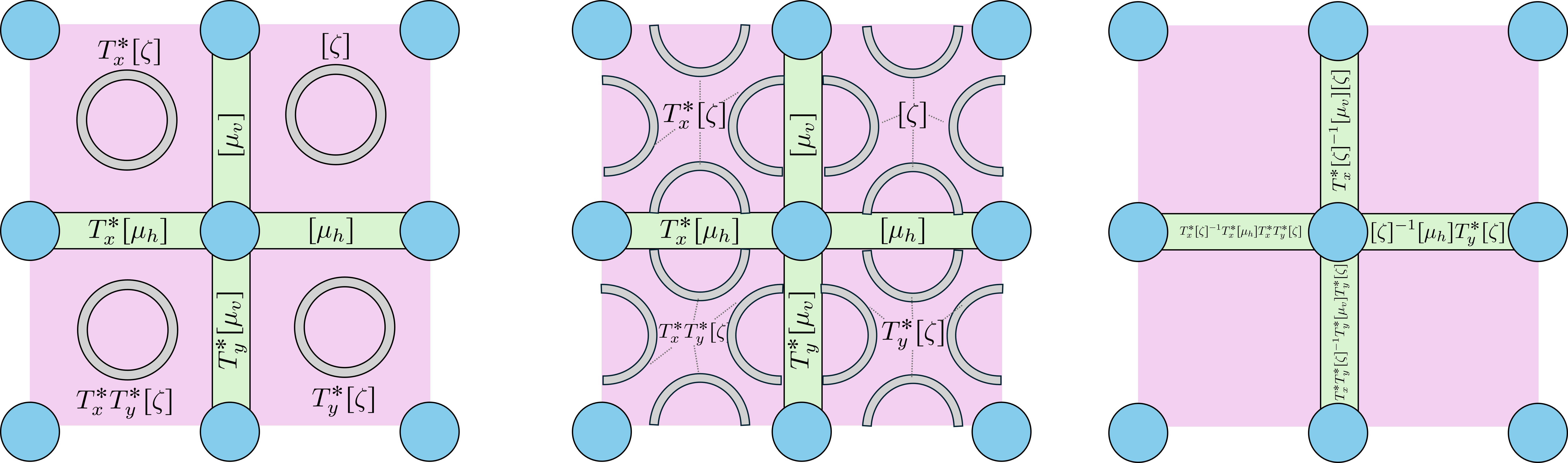}
    \caption{Physical process that produces the equivalence Eq.~\ref{eqn:1CellEquivalence}. (1+1)D SPTs (grey circles) are created in a translation-invariant way on each 2-cell (pink), then expanded and merged into the 1-cells (green). The inverses arise from tracking the orientations of the 1-cells and the (1+1)D SPTs.}
    \label{fig:1CellEquivalence}
\end{figure}

Moving to (2+1)D, again with $G_{sp}$ consisting only of translations, the horizontal and vertical 1-cell weak SPT data $[\mu_v]$ and $[\mu_h]$ each naively form a $\H^2(G_{int},\U1)$ torsor. We find an equivalence
\begin{equation}
    ([\mu_v],[\mu_h]) \sim \left(\frac{T_x^\ast [\zeta]}{[\zeta]}[\mu_v],\frac{[\zeta]}{T_y^\ast[\zeta]}[\mu_h]\right)
    \label{eqn:1CellEquivalence}
\end{equation}
for any $[\zeta] \in \H^2(G_{int},\U1)$
This equivalence is obtained by nucleating the (1+1)D $G_{int}$-SPT $[\zeta]$ in the center of a reference 2-cell, and then expanding that SPT out to the edges, as shown in Fig.~\ref{fig:1CellEquivalence}. This an again be done in a $G$-symmetric way, provided we account for the action of $T^\ast$ to preserve translation symmetry. This equivalence relation is trivial for an unmodulated symmetry since any two neighboring 2-cells contribute opposite SPTs to their shared 1-cell.

\subsection{Classification of (1+1)D MSPTs with only internal and translation symmetries}

The above discussion leads to an immediate classification of (1+1)D MSPTs with $G=G_{int}\rtimes G_{sp}$ with $G_{sp}=\Z$, where $\Z$ represents translations. The data on the 1-cells is classified by $T^\ast$-invariant elements of $\H^2(G_{int},\U1)$, and as discussed in Eq.~\ref{eqn:0CellEquivalence}, the data on the 0-cells forms a torsor over coinvariants of $\H^1(G_{int},\U1)$. That is, distinct MSPTs form a group
\begin{equation}
    \H^2(G_{int},\U1) \times \frac{\H^1(G_{int},\U1)}{(T^\ast-1)\H^1(G_{int},\U1)} = \H^2(G_{int}\rtimes G_{sp},\U1)
\end{equation}
where the last equality follows from the Lyndon-Hochschild-Serre (LHS) spectral sequence. This result verifies the crystalline equivalence principle for modulated symmetries of this form, giving evidence that the crystalline equivalence principle works for modulated symmetries as well as ordinary symmetries.

\subsection{Extended symmetry on cells}
\label{subsec:extendedSymm}

We assumed in our earlier derivations that no lower-dimensional cells are invariant under an element of $G_{sp}$. Let us relax that assumption. Suppose, for example, that a $(d-1)$-cell is invariant under $G_{\Sigma_{d-1}}$. Then we need to construct a gapped interface between our two $d$-cells which is invariant under the extended (internal) symmetry $G_m \rtimes G_{\Sigma_{d-1}}$ rather than just $G_m$.

Note that characterizing such a gapped interface is \textit{not} covered by a naive application of the results of Ref.~\cite{wang2018symmetric}. Applying those results would require the ``extra" symmetry $G_{\Sigma_{d-1}}$ to be a normal subgroup of the extended symmetry, which is not true in our case; instead $G_m$ is a normal subgroup of the extended symmetry. We argue that the anomaly condition Eq.~\ref{eqn:modulatedAnomaly} is still correct (although the argument is more technically involved than the one we gave previously), but the classification of weak SPT data is modified by the presence of the additional symmetry. We will give the argument for $(d-1)$-cells; the lower-dimensional cells follow analogously.

For future use, we quote a technical result arising from the Lyndon-Hochschild-Serre (LHS) spectral sequence~\cite{WangDomainWalls} and use it in an argument inspired by Ref.~\cite{MixedStateSPT}. For a symmetry group $K = G_{int} \rtimes G_{\Sigma_{d-1}}$, the LHS spectral sequence degenerates at the $E_2$ page and shows that $\H^{d+1}(K,\U1)$ includes a subgroup
\begin{equation}
    \H^0(G_{\Sigma_{d-1}},\H^{d+1}(G_{int},\U1)) = \H^{d+1}(G_{int},\U1)^S \equiv \left\lbrace [\omega] \in \H^{d+1}(G_{int},\U1) \big| S^\ast [\omega] = [\omega] \text{ for all } S \in G_{\sigma_{d-1}}\right\rbrace
    \label{eqn:spectralSequenceGInvariant}
\end{equation}
where the $\H^{d+1}(G_{int},\U1)$ coefficients are twisted by the action of $G_{\Sigma_{d-1}}$. This subgroup can be physically interpreted as $G_{\Sigma_{d-1}}$-symmetric $G_{int}$-SPTs.

We will give the argument for $G_{\Sigma_{d-1}}=\Z_2$ where the nontrivial element of $\Z_2$ is a reflection $R$; the generalization to other symmetries is conceptually identical.

\begin{figure}
    \centering
    \begin{subfigure}[b]{0.5\textwidth}
    \includegraphics[width=\textwidth]{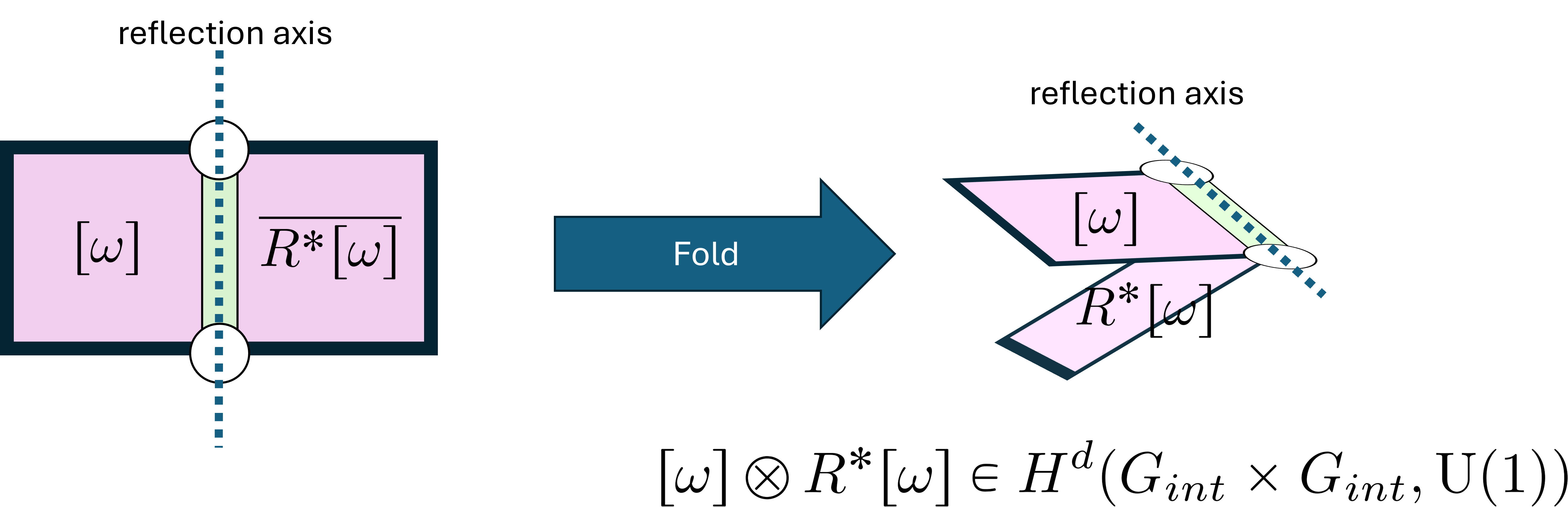}
    \caption{}
    \label{fig:foldingTrick}
    \end{subfigure}
    \begin{subfigure}[b]{0.6\textwidth}
        \includegraphics[width=\textwidth]{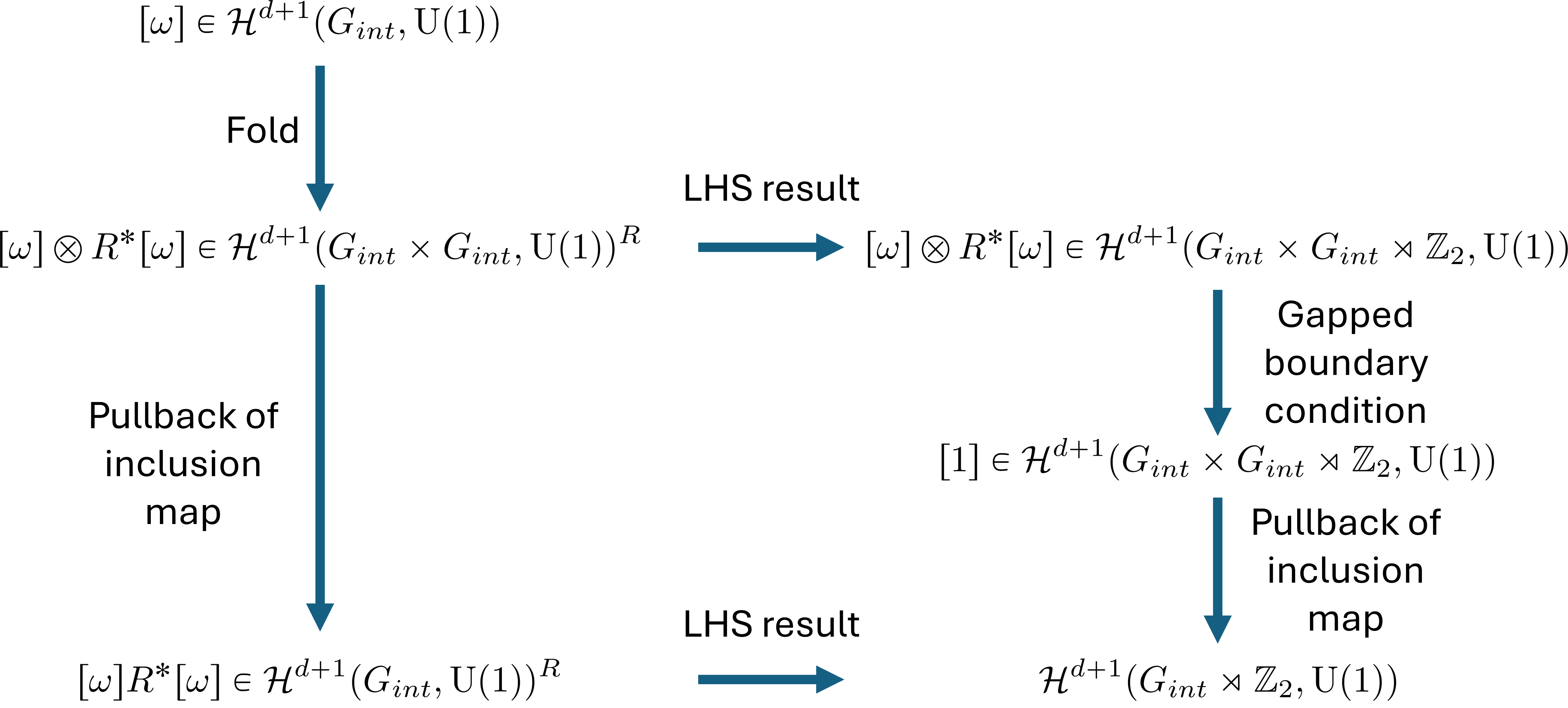}
        \caption{}
        \label{fig:commutativeDiagram}
    \end{subfigure}
    \caption{(a) Folding trick used to take a gapped interface between two $d$-cells transforming under $G_{int}\times G_{int}$ with reflection symmetry into a single (double-layered) SPT transforming under the same symmetry group. Reflection preserves the boundary and swaps the folded copies. (b) Commutative diagram used to argue for the anomaly-free condition Eq.~\ref{eqn:modulatedAnomaly} in the presence of reflection symmetry. The $R$ superscript means the $R^\ast$-invariant subgroup.}
\end{figure}

Consider a $(d-1)$-cell which is left invariant under $R$. If the data on the $d$-cell on one side of the $(d-1)$-cell is $\omega$, then $\overline{R^\ast\omega}$ appears on the other side, where we are tracking the fact that $R$ reverses orientation by defining $R^\ast$ to have a trivial action on the $\U1$ coefficients and manually inserting the complex conjugation arising from the orientation reversal. Now fold the system on the $(d-1)$-cell, as shown in Fig.~\ref{fig:foldingTrick}. Then the two $d$-cells form a $G_{int}\times G_{int}$ SPT given by the class $\tilde{\omega} = \omega \otimes R^\ast \omega$. The action of $R$ on this enlarged symmetry group swaps the two copies of $G_{int}$ (and does not complex conjugate the cohomology class, since the two folded copies have opposite orientation), so $\tilde{\omega}$ is symmetric under $R^\ast$.

We now make an argument for Eq.~\ref{eqn:modulatedAnomaly} that is outlined in the commutative diagram in Fig.~\ref{fig:commutativeDiagram}.

The boundary should break the $G_{int}\times G_{int}\rtimes G_{\Sigma_{d-1}}$ symmetry down to a $G_{int}\rtimes G_{\Sigma_{d-1}}$ symmetry, and we are looking for a gapped interface. Since we have already accounted for the group action of $G_{\Sigma_{d-1}}$ on $\omega$, we are breaking the symmetry down to the \textit{diagonal} subgroup of $G_{int} \times G_{int}$. (In the previous section, we instead worked with $\omega \otimes \overline{\omega}$ and broke the symmetry down to the non-diagonal subgroup. For our present purposes it is simpler to anticipate the result and take this equivalent perspective.) 

To have a gapped interface, then, we require that some cohomology class is trivial in $\H^{d+1}(G_{int}\rtimes G_{\Sigma_{d-1}},\U1)$ is trivial. But which class? The standard theory of symmetry-breaking SPT boundaries~\cite{wang2018symmetric} tells us that we should view the boundary symmetry $G_{int}\rtimes G_{\Sigma_{d-1}}$ as a subgroup of the bulk symmetry $G_{int} \times G_{int} \rtimes G_{\Sigma_{d-1}}$, and pull back the bulk $G_{int} \times G_{int} \rtimes G_{\Sigma_{d-1}}$ cocycle to a boundary $G_{int} \rtimes G_{\Sigma_{d-1}}$ cocycle. If this pullback is trivial, then there is a gapped interface. Using the spectral sequence results in Eq.~\ref{eqn:spectralSequenceGInvariant}, we can view $\tilde{\omega}$ as a $G_{int} \times G_{int} \rtimes G_{\Sigma_{d-1}}$ cocycle, which can then be pulled back to $G_{int} \rtimes G_{\Sigma_{d-1}}$. This pulled-back version of $\tilde{\omega}$ must be trivial to have a gapped interface.

Equivalently, we can do the pullback and spectral sequence   inclusion in the opposite order and obtain the same cocycle. First pull back $\tilde{\omega}$ as a $G_{int} \times G_{int}$ cocycle to a $G_{int}$ cocycle, which is clearly $R^\ast$-invariant and equal to $\omega R^\ast \omega$, with no tensor product. Then view the pulled-back cocycle as an element of $\H^{d+1}(G_{int} \rtimes G_{\Sigma_{d-1}},\U1)$. We can therefore conclude that the gapped interface condition is
\begin{equation}
    [\omega] R^\ast [\omega]=1
\end{equation}
which (accounting for the complex conjugation in the un-folded picture) is the same anomaly-free condition we had previously in Eq.~\ref{eqn:modulatedAnomaly}.

The above discussion makes it clear that naively distinct weak SPT data on the $(d-1)$-cells is given by distinct trivializations of $\omega R^\ast \omega$ as an element of $Z^{d+1}(G_{int} \rtimes G_{\Sigma_{d-1}},\U1)$, which form an $\H^d(G_{int} \rtimes G_{\Sigma_{d-1}},\U1)$ torsor in the standard way. 

There is a subtlety in the above - if a reflection is in $G_{\Sigma_{d-1}}$, it does \textit{not} reverse the orientation of the $(d-1)$-cell it leaves fixed (by our Wyckoff assumptions above). Therefore, the reflection should act \textit{unitarily} on the $\U1$ coefficients. This follows from our arguments above, since after the folding occurs, reflection now acts unitarily on the coefficients. Note that although the extended symmetry does not change the anomaly-free condition, it does change (in general) the classification of the weak SPT data. In general, the equivalence relations on the naive weak SPT data are nontrivial and depend on the situation. We will consider a (1+1)D example in Sec.~\ref{sec:1DReflection}.

\section{A simple example in (1+1)D - the cluster state}
    \label{sec:1DCluster}

    As a simple example, consider the $\Z_N$ cluster model in (1+1)D. The exactly solvable model for this SPT consists of $\Z_N$ qudits on the sites of a lattice, with Hamiltonian
    \begin{equation}
        H = -\sum_{i \text{even}} Z_{i-1}^kX_iZ_{i+1}^{-k} - \sum_{i \text{odd}} Z^{-k}_{i-1}X_iZ_{i+1}^k + \text{ h.c.}
    \end{equation}
    where $X_i$ and $Z_i$ are the usual clock and phase operators for $\Z_N$ qudits, and $k$ is a $\Z_N$ parameter. This Hamiltonian has two internal symmetries,
    \begin{equation}
        Q_1 = \prod_{i}X_{2i} \quad \text{and} \quad Q_2 = \prod_i X_{2i+1},
    \end{equation}
    which states that the total $\Z_N$ charge on the even and odd sublattices are conserved. We also see that translation symmetry $T$ by a single site does not commute with $H$ unless $k=0$ or $N$ is even and $k=N/2$ (since $Z^N = 1$). It is generally known~\cite{geraedts2014SPTs,Santos2015SPTs} that this model realizes the $\Z_N \times \Z_N$ SPT given by the class $[k] \in \H^2(\Z_N \times \Z_N,\U1)=\Z_N$. We also see explicitly that translation $T$ by a single site acts as
    \begin{equation}
        TQ_{1,2}T^{-1} = Q_{2,1}
        \label{eqn:translationSwap}
    \end{equation}
    or, in our notation, translation has a group action $T(Q_{1,2})=Q_{2,1}$.
    
    A result in Ref.~\cite{han2023modulated} demonstrated that the interplay between $k$, translation symmetry, and the modulated symmetries in this behavior is generic. Namely, when translation interchanges the two $\Z_N$ symmetries via. Eq.~\ref{eqn:translationSwap}, the only translation-symmetric $\Z_N \times \Z_N$ SPTs are $[0]$ and, for $N$ even, $[N/2]$. We will use our formalism to reproduce this result, but as we will see in the following sections, the same technique can be used in a much more general setting.

    \begin{figure}
        \centering
        \includegraphics[width=0.5\linewidth]{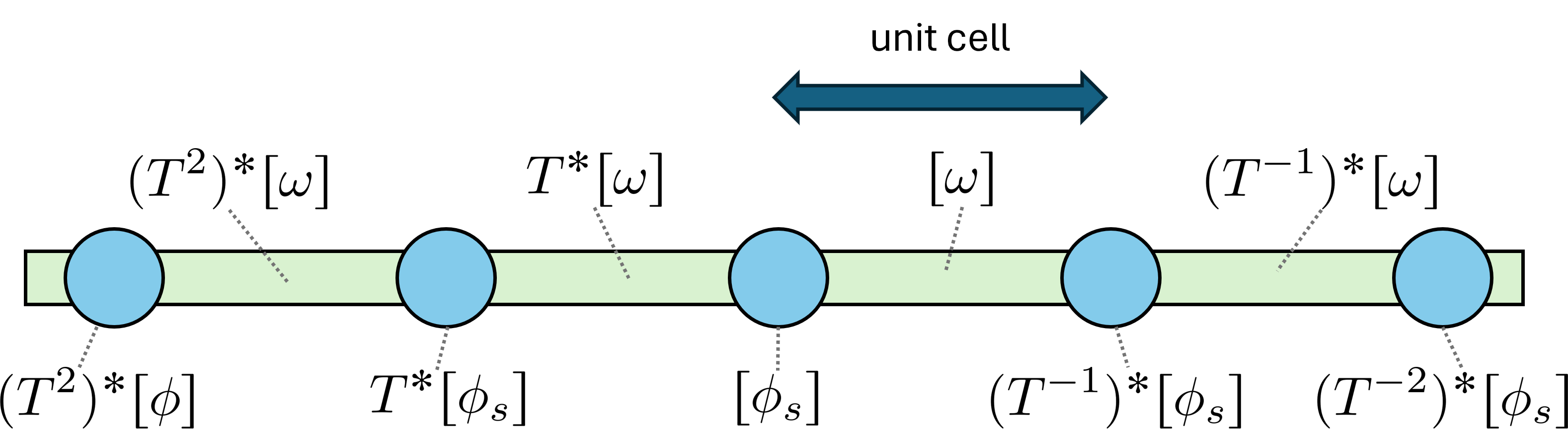}
        \caption{Defect network for a (1+1)D state with translation symmetry only. 1-cells are green bars, 0-cells are blue circles. The data $[\omega] \in \H^2(G_{int},\U1)$ is assigned to a reference 1-cell, and then propagated to other 1-cells via translation symmetry. The data $[\phi_s] \in C^1(G_{int},\U1)/B^1(G_{int},\U1)$ is assigned to a reference 0-cell and propagated to other 0-cells via translation symmetry.}
        \label{fig:ClusterNetwork}
    \end{figure}

    The cellulation we use and the assignment of data to the cells are shown in Fig.~\ref{fig:ClusterNetwork}. For this case, we choose a $[\omega] = [k] \in \H^2(\Z_N \times \Z_N, \U1)$ to place on a representative 1-cell. A 2-cocycle representative $\omega$ for $[k]$ is given by
    \begin{equation}
        \omega(g,h) = e^{\frac{2 \pi i k}{N}g_1h_2}
        \label{eqn:ZnZnCocycle}
    \end{equation}
    where $g = (g_1,g_2) \in \Z_N \times \Z_N$. Then
    \begin{equation}
        T^\ast\omega(g,h) = e^{\frac{2\pi i k}{N} h_1 g_2}.
    \end{equation}
    Multiply this by the coboundary generated by the 1-cochain
    \begin{equation}
        f(g) = e^{\frac{2\pi i k}{N} g_1 g_2},
    \end{equation}
    we obtain
    \begin{align}
         T^\ast\omega df &= e^{-\frac{2\pi i k}{N} g_1 h_2} = \overline{\omega}
    \end{align}
    We therefore see that the anomaly-free condition is
    \begin{equation}
        [k] = T^\ast\left(\left[k\right]\right)=\left[-k\right]
        \label{eqn:ClusterAnomaly}
    \end{equation}
    Hence, the vanishing of the anomaly forces $2k=0\mod N$, i.e., $k=0$ if $N$ is odd or $k=0,N/2$ if $N$ is even. This reproduces our desired result.

    In this case, the weak SPT data placed on the 0-cells is  naively classified by $\H^1(\Z_N \times \Z_N,\U1) = \Z_N^2$ and has the interpretation of an arbitrary charge per unit cell under each $\Z_N$ symmetry. Notably, this data need \textit{not} be invariant under $T^\ast$ (in fact, only a $\Z_N$ subgroup of $\Z_N^2$ is translation-invariant!); what matters is that we only have a free choice of an element of $\H^1(\Z_N \times \Z_N,\U1)$ on a single site, and the data on the rest of the sites are generated by translation.

    Following Eq.~\ref{eqn:0CellEquivalence}, there is an equivalence relation on the weak SPT data $(a,b) \in \Z_N^2$ of the form
    \begin{equation}
        (a,b) \sim (b,a) \quad \Rightarrow \quad (x,-x) \sim (0,0)
    \end{equation}
    We therefore need to mod out the antisymmetric $\Z_N$ subgroup of the weak SPT data.

    The overall result, then, is that there is a strong SPT classification of $\Z_{(2,N)}$ and a weak classification of $\Z_N$. This matches the crystalline equivalence principle expectation where the overall classification should be
    \begin{equation}
        \H^2(\Z_N \times \Z_N \rtimes_{\mathrm{cluster}} \Z, \U1) = \Z_N \times \Z_{(2,N)}
    \end{equation}

\section{Classification of dipolar SPTs in (1+1)D}
    \label{sec:1DDipolar}

    We will now use our formalism to classify SPTs in (1+1)D protected by various types of dipolar symmetries, namely $\Z_N$, $\U1$, and generic finite Abelian group dipolar symmetries. The classification is new for $\U1$ dipolar symmetries, and we reproduce the known strong SPT classification for $\Z_N$ and finite Abelian group dipolar symmmetries~\cite{han2023modulated,lam2023dipole}. In the known cases, we also obtain the weak SPT data. We will also connect the previously obtained classification of dipolar SPT phases via matrix product states~\cite{lam2023dipole} to our formalism and reinterpret those results. In all cases, the defect network is set up as in Fig.~\ref{fig:ClusterNetwork}, but $G_{int}$ and the action of $T$ on $G_{int}$ varies depending on the example.

    \subsection{\texorpdfstring{$\Z_N$}{ZN} dipolar symmetry}
    \label{sec:1DZnDipolar}

    We first use our formalism to show that there is a $\Z_N$ classification of strong SPTs protected by $\Z_N^{(Q)} \times Z_N^{(x)}$ symmetry (where $Q$ refers to charge conservation and $x$ refers to dipole conservation in the $x$ direction) with translation, and then obtain $\Z_N$ weak SPT indices.
    
    A dipolar symmetry obeys the following algebra with translation,
    \begin{align}
        T(Q) &= Q\\
        T(D) &= QD,
    \end{align}
    where $Q$ is the charge conservation symmetry and $D$ is the dipolar symmetry. For this section, we assume $Q^N=D^N=1$. Note that we can choose a convention where $T(D)$ is either $QD$ or $Q^{-1}$D; our results are independent of this choice, which physically corresponds to choosing active versus passive translations.
    
    We notate $g = (g_Q,g_x) \in \Z_N^{(Q)}\times \Z_N^{(x)}$, where $g_Q$ and $g_x$ take additive $\Z_N$ values. If we pick the 1-cell data 
    \begin{equation}
        [\omega] = [k] \in \H^2(\Z_N \times \Z_N,\U1) = \Z_N
    \end{equation}
    and use the representative cocycle in Eq.~\eqref{eqn:ZnZnCocycle}, then we compute that
    \begin{equation}
        T^\ast(\omega)(g,h) = e^{\frac{2\pi i k}{N} [g_Q+g_x]h_x}
    \end{equation}
    where the brackets indicate addition modulo $N$. It is straightforward to check that the brackets can be safely dropped.
    
    Now multiply by a coboundary generated by the 1-cochain
    \begin{equation}
        f(g) = \begin{cases} 
                    e^{\frac{2\pi i k}{2N} g_x^2} & N \text{ even}\\
                    e^{2\pi i k\frac{N+1}{2N}g_x^2} & N \text{ odd}
                \end{cases}
    \end{equation}
    which is straightforwardly checked to be single-valued for $g_Q \in \Z_N$. Regardless of the parity of $N$,
    \begin{equation}
        df(g,h) = e^{-\frac{2\pi i k}{N} g_x h_x}
    \end{equation}
    so that
    \begin{equation}
        \left(T^\ast\omega\right)df = \omega.
    \end{equation}
    This shows that $T^\ast\left[\omega\right] = \left[\omega\right]$, so we may use any $[\omega]$ in constructing a dipole SPT. Hence the strong SPT classification is $\Z_N$.
    
    \subsubsection{Weak dipolar SPT data}
    \label{sec:1DDipole0Cell}
    We can refine the classification by adding weak SPT data, which amounts to dressing the 0-cells by (0+1)D SPTs. Since the symmetry which preserves the 0-cells is still $\Z_N \times \Z_N$, the $(d-1)$-cell data naively form an $\H^1(\Z_N \times \Z_N, \U1) = \Z_N \times \Z_N$ torsor.
    
    To see that translation symmetry is required to distinguish these weak SPT classes, consider two choices of 0-cell data that differ by $[k] \in \H^1(\Z_N \times \Z_N, \U1)$ (using additive notation). In the absence of translation symmetry, on every $N$th 0-cell, we can create $N$ copies of the $[k]$ $(0+1)$D $\Z_N \times \Z_N$ SPT with a constant depth circuit because $[Nk]=[0]$. Then we can move those copies so that each 0-cell contains a copy, transforming the original 0-cell data into the second one. Hence, all 0-cell data is equivalent if translation symmetry is broken.

    Applying Eq.~\ref{eqn:0CellEquivalence}, we check that if $(k_Q,k_x) \in \H1^(\Z_N \times \Z_N,\U1)$, then for dipolar symmetry
    \begin{equation}
        T^\ast(k_Q,k_x) = (k_Q,k_x+k_Q) \sim (k_Q,k_x) \quad \Rightarrow \quad  (0,a) \sim (0,0)
        \label{eqn:1DDipoleEquivalence}
    \end{equation}
    Hence we need to mod out the pure dipolar weak SPT data, reducing the weak classification to $\Z_N$. The overall classification of (1+1)D dipole SPTs therefore contains $\Z_N$ strong SPTs and $\Z_N$ weak SPTs. One can check that this is consistent with the crystalline equivalence principle.
    
\begin{figure}
        \centering
        \includegraphics[width=0.5\linewidth]{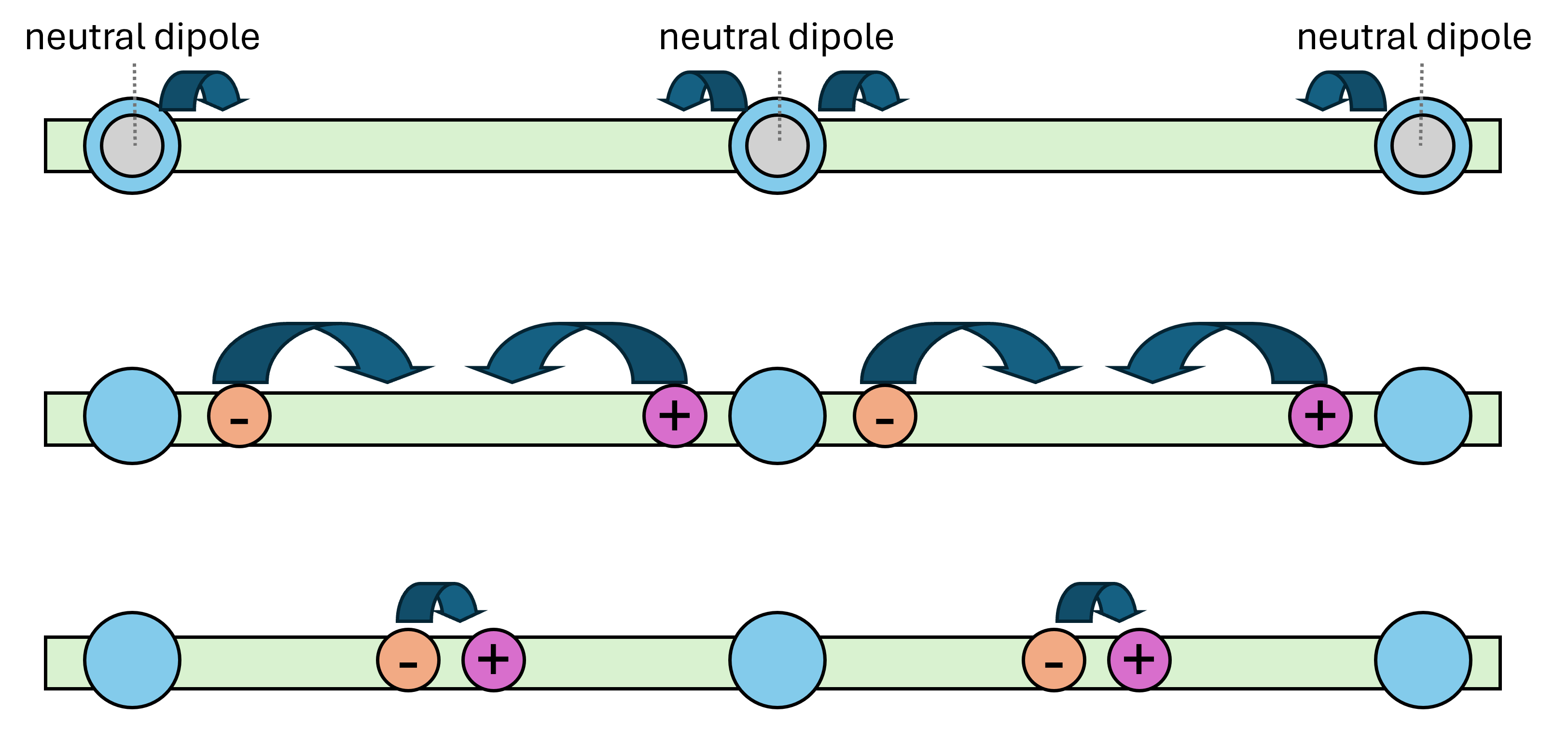}
        \caption{Triviality of the weak SPT with a neutral dipole moment per unit cell. The neutral dipole (grey) on the 0-cells (blue) can be symmetrically separated into positive (orange) and negative (purple) charges, then annihilated on the center of a 1-cell (green).}
        \label{fig:trivialNeutralDipole}
    \end{figure}

    Let us physically interpret the 0-cell data for the weak SPTs. We temporarily ignore the equivalence Eq.~\ref{eqn:1DDipoleEquivalence}. For simplicity, let us choose the trivial strong SPT with $\omega(g,h) = 1$. Then the 0-cell data is itself an element $[(k_Q,k_x)] \in \Z_N \times \Z_N$ with the representative cocycle
    \begin{equation}
        (k_Q,k_x)(g) = e^{\frac{2\pi i}{N}(k_Q g_Q+k_xg_x)}
    \end{equation}
    Under translation,
    \begin{equation}
        T^\ast[(k_Q,k_x)] = [(k_Q,k_x+k_Q)]
    \end{equation}
    It is natural to think of $k_Q$ as labeling the charge per unit cell, i.e., the $\Z_N$ filling fraction. Due to the action of translation, $k_x$ is a little more subtle than the dipole moment per unit cell. In particular, on site 0, suppose we place $(k_Q,k_x)$. Then on site $j$, we have $(k_Q,k_x+jk_Q)$. Hence the total dipole charge $Q_d$ (which is only defined modulo $N$) is
    \begin{equation}
        Q_d= \sum_j (k_x + jk_Q) \mod N
    \end{equation}
    We therefore should interpret $k_x$ as a charge-neutral $\Z_N$ dipole moment per 0-cell. Such a dipole moment is trivial; we can split it into a positive charge on the left of the 0-cell and a negative charge on the right of the 0-cell, and then combine these charges in the center of the 1-cell, as shown in Fig.~\ref{fig:trivialNeutralDipole}. This triviality is the reason for the equivalence Eq.~\ref{eqn:1DDipoleEquivalence}.

    \subsection{\texorpdfstring{$\U1$}{U(1)} dipolar symmetry}
        \label{subsec:U1dipolar1D}
        
        Our symmetry group is $\U1^2$, with charge and dipole symmetry generating each copy. A straightforward application of the K\"unneth formula tells us that
        \begin{equation}
            \H^2(\U1^2,\U1) = 0.
        \end{equation}
        Hence there are no strong $\U1$ dipolar SPTs in (1+1)D. The 0-cell data is given by an element of
        \begin{equation}
           \H^1(\U1^2,\U1) = \Z \times \Z
        \end{equation}
        which corresponds to weak SPTs with a charge or a dipole per unit cell, much like in the $\Z_N$ case. Again, due to Eq.~\ref{eqn:1DDipoleEquivalence}, the pure dipolar weak SPT is modded out, so the overall classification is $\Z$.

    \subsection{Generic finite Abelian dipolar symmetry}

    Suppose that we have a system whose charge is valued in an Abelian group $G$, and suppose that we impose dipolar $G$ symmetry. By the fundamental theorem of Abelian groups, we may write
    \begin{equation}
        G = \prod_{\ell} \Z_{n_\ell}
    \end{equation}
    where $n_\ell$ are integers and $\ell$ runs over some finite list. Then the internal symmetry is $G_{int}=G \times G$, with group elements labeled by $(g_{\ell,Q},g_{\ell,D})$ for each $\ell$, where $Q$ and $D$ indicate the charge and dipolar parts of the $\Z_{n_\ell}$ charge. Here $g_{\ell,Q}$ and $g_{\ell,D}$ are both valued in $\Z_{n_\ell}$.

    Translation acts as
    \begin{equation}
        T(g_{\ell,Q},g_{\ell,D}) = (g_{\ell,Q}+g_{\ell,D},g_{\ell,D})
    \end{equation}
    Explicit representative cocycles for $\H^2(G_{int},\U1)$ take the form
    \begin{align}
        \omega_{iQ,jQ}(g,h) &= e^{\frac{2\pi i}{\gcd(n_i,n_j)}k_{iQ,jQ}g_{i,Q}h_{j,Q}}\\
        \omega_{iD,jD}(g,h) &= e^{\frac{2\pi i}{\gcd(n_i,n_j)}k_{iD,jD}g_{i,D}h_{j,D}}\\
        \omega_{iQ,jD}(g,h) &= e^{\frac{2\pi i}{\gcd(n_i,n_j)}k_{iQ,jD}g_{i,Q}h_{j,D}}
    \end{align}
    Here $i<j$ for the first two lines and there is no constraint on $i$ and $j$ in the last line. 
    The first two lines each generate a copy of $\H^2(G,\U1)$ and the last line generates $\H^1(G,\H^1(G,\U1))$. By explicit computation,
    \begin{align}
        T^\ast[\omega_{iQ,jQ}] &= [\omega_{iQ,jQ}][\omega_{iQ,jD}][\omega_{jQ,iD}]^{-1}[\omega_{iD,jD}]\\
        T^\ast[\omega_{iD,jD}] &= [\omega_{iD,jD}]\\
        T^\ast[\omega_{iQ,jD}] &= [\omega_{iQ,jD}][\omega_{iD,jD}]
    \end{align}
    where in the last line, if $i=j$, $[\omega_{iD,iD}]$ is trivial. Hence the only translation-invariant cocycles are the copy of $\H^2(G,\U1)$ generated by $[\omega_{iD,jD}]$ and the subgroup of $\H^1(G,\H^1(G,\U1)$ that is symmetric in $i$ and $j$. One can check that the antisymmetric part of $\H^1(G,\H^1(G,\U1)$ is isomorphic to $\H^2(G,\U1)$, so we conclude that the strong SPT data is classified by
    \begin{equation}
        \H^2(G,\U1) \times \H^1(G,\H^1(G,\U1)^{\text{symm}} \sim \H^1(G,\H^1(G,\U1)
    \end{equation}
    where the twiddle means that the groups are isomorphic.
    As a group, this matches the result of Ref.~\cite{lam2023dipole}, which gives a classification of
    \begin{equation}
        \H^2(G\times G,\U1)/\left(\H^2(G,\U1)\right)^2 \sim \H^1(G,\H^1(G,\U1)).
        \label{eqn:lamClassification}
    \end{equation}
    where the last equality follows from applying the K\"unneth formula to the numerator.

    As before, the weak SPT data will naively form a $\H^1(G \times G,\U1) = \left(\H^1(G,\U1)\right)^2$ torsor. The equivalence Eq.~\ref{eqn:0CellEquivalence} again removes the pure dipolar weak SPTs, so the weak SPT data is just one copy of $\H^1(G,\U1)$.

    \subsection{Lam's classification of dipolar SPTs}
        We will now summarize the result in Ref.~\cite{lam2023dipole} computing the SPT classification for dipolar SPTs with a generic Abelian charge $G$ in a way which connects directly to our language from the previous section and show explicitly that quotient by copies of $\H^2(G,\U1)$ can be thought of as a constraint arising from Eq.~\ref{eqn:modulatedAnomaly} rather than a quotient. 
        
        Ref.~\cite{lam2023dipole} uses matrix product state (MPS) language to construct the edge operators for a dipole SPT on a finite size chain. Specifically, Ref.~\cite{lam2023dipole} shows the following. Suppose the MPS is constructed from a matrix $A_{ij}^m$, where $m$ is the physical index and $i$ and $j$ label virtual indices, and the symmetry operators are
        \begin{align}
            Q_g &= \bigotimes_{\text{sites } n} U_g\\
            D_g &= \bigotimes_{\text{sites } n} \left(U_g\right)^n.
        \end{align}
        Translation symmetry is implicitly assumed in  Ref.~\cite{lam2023dipole} (except for the fact that the chain is finite size), so we have a true modulated symmetry.
        
        Invariance under the charge operators forces there to be a matrix $V_g$ such that $U_g$ can be pushed to the virtual indices as
        \begin{equation}
           \sum_\ell \left(U_g\right)^{m\ell}A_{ij}^\ell = e^{i\theta_g}\left(V_g A^m V_g^\dagger\right)_{ij}
        \end{equation}
        where the matrix multiplication on the right hand side acts on virtual indices only. Likewise, Ref.~\cite{lam2023dipole} shows that there exists a matrix that pushes copies of $V$ from one virtual index to the other:
        \begin{equation}
            (V_gA^m)_{ij} = e^{i\vartheta_g}(W_gA^mW_g^\dagger)_{ij}
        \end{equation}
        One can then show with MPS graphical calculus that on a finite size system, the charge and dipole operators can be pushed through to the dangling virtual indices such that
        \begin{align}
            Q_g &= Q_g^LQ_g^R\\
            D_g &= D_g^L D_g^R
        \end{align}
        where the operators $Q_g$, $D_g$ act on real indices, $Q_g^L$ and $D_g^L$ act on the dangling virtual index on the left, and $Q_g^R$ and $D_g^R$ act on the dangling virtual index on the right. The operators are
        \begin{align}
            Q_g^L &= V_g\\
            Q_g^R &= V_g^\dagger\\
            D_g^L &= W_g\\
            D_g^R &= (V_g^\dagger)^{L_0}W_g^\dagger
        \end{align}
        where the system size is $L_0$ sites. Now, the commutation relations of the operators $Q_g^L$ and $D_g^L$ can be computed and will generate a cocycle $\omega^L \in \H^2(G \times G,\U1)$, in that $Q_g^L$ and $D_g^L$ can form a projective representation of $G \times G$. The operators $\left(Q_g^R\right)^\dagger$ and $\left(D_g^R\right)^\dagger$ also generate some projective representation $\omega^R$ of $G \times G$. In order for $Q_g$ and $D_g$ to form a linear representation of $G \times G$, we must have $\omega^L = \omega^R$ as cohomology classes for any system size $L_0$.
        
        Observe further that since $T(D_g) = Q_gD_g$, we have $(T^\ast)^{L_0}(\omega^L) = \omega^R$. Hence, demanding $\omega^L = \omega^R$ for all $L_0$ is the same as requiring
        \begin{equation}
            T^\ast \omega^L = \omega^L \text{ for } L_0=1
            \label{eqn:lamAnomalyFree}
        \end{equation}
        where we are formally taking the operator algebra defined by $L_0=1$. That is, dipolar SPTs are specified by a 2-cocycle $\omega \in \H^2(G\times G,\U1)$ such that $T^\ast \omega = \omega$. This rephrases the results of Ref.~\cite{lam2023dipole} into our language.
        
        Notice that this is a \textit{constraint} on $\omega$. It is not saying that we should identify SPTs which do not obey Eq.~\ref{eqn:lamAnomalyFree} as trivial, i.e., quotient out by such SPTs; it is saying that we should exclude all the ones which do not satisfy the constraint. Mathematically speaking, the result does not depend on whether we quotient or exclude; for finite Abelian $G$ all the cohomology groups involved are also finite Abelian, so the operations of taking the non-anomalous subgroup and quotienting out the anomalous subgroup will agree. From a physical point of view, it is better to view the two copies of $\H^2(G,\U1)$ which appear in the denominator of Eq.~\ref{eqn:lamClassification} as an anomalous subgroup rather than as a set of trivial phases.
        
        We note further that according to the K\"unneth formula,
        \begin{align}
            \H^2(G\times G,\U1) &= \H^2(G,\U1) \oplus \H^1(G,\H^1(G,\U1)) \oplus \H^0(G,\H^2(G,\U1))\\
            &=\H^2(G,\U1)^2 \oplus \H^1(G,\H^1(G,\U1))
        \end{align}
        Therefore, removing the $\H^2(G,\U1)^2$ subgroup leaves behind a strong SPT classification of
        \begin{equation}
            \H^1(G,\H^1(G,\U1)).
        \end{equation}
        
        Ref.~\cite{lam2023dipole} actually writes down an explicit ansatz for $\omega$, but as a set of projective representations for each copy of $G$ (i.e. it separates the algebra of the $V_g$ and the algebra of the $W_g$). This ansatz is not gauge-invariant, so we will not attempt to reproduce that language.

\section{Other results in (1+1)D}
    \label{sec:Other1D}

    Our construction allows us to use a single formalism to reproduce a number of known results in (1+1)D which have been derived from many different considerations, as well as to obtain new results.

    \subsection{(1+1)D cluster state with reflection symmetry}
    \label{sec:1DReflection}
    
    In order to see what happens when we add point group symmetries in the presence of modulated symmetries, we can consider the symmetries of the (1+1)D cluster state in Sec.~\ref{sec:1DCluster}, but impose an additional reflection symmetry $R$. For simplicity we choose our reflection axis to live on a site, which commutes with $Q_1$ and $Q_2$. One could also compose with a translation by an odd number of sites to consider a reflection about a bond center; this just changes the presentation of the space group and leads to the same results.

    \begin{figure}
        \begin{subfigure}[b]{0.55\textwidth}
            \includegraphics[width=\textwidth]{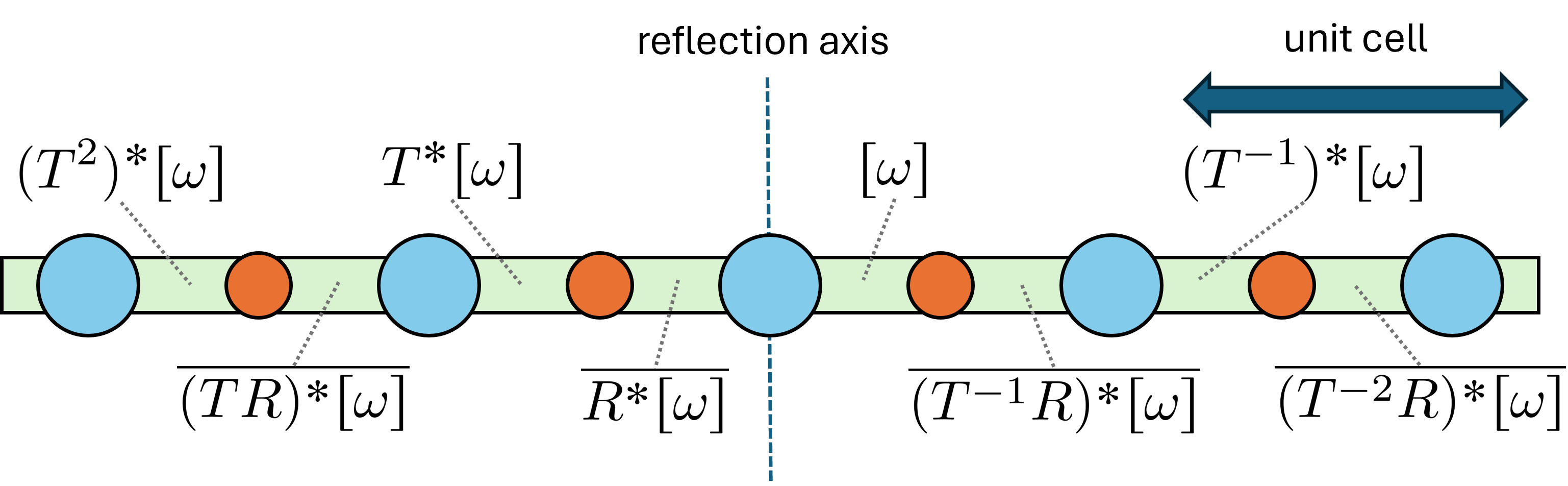}
            \caption{}
            \label{fig:cellulation1DReflection1Cell}
        \end{subfigure}
        \hfill
        \begin{subfigure}[b]{0.55\textwidth}
            \includegraphics[width=\textwidth]{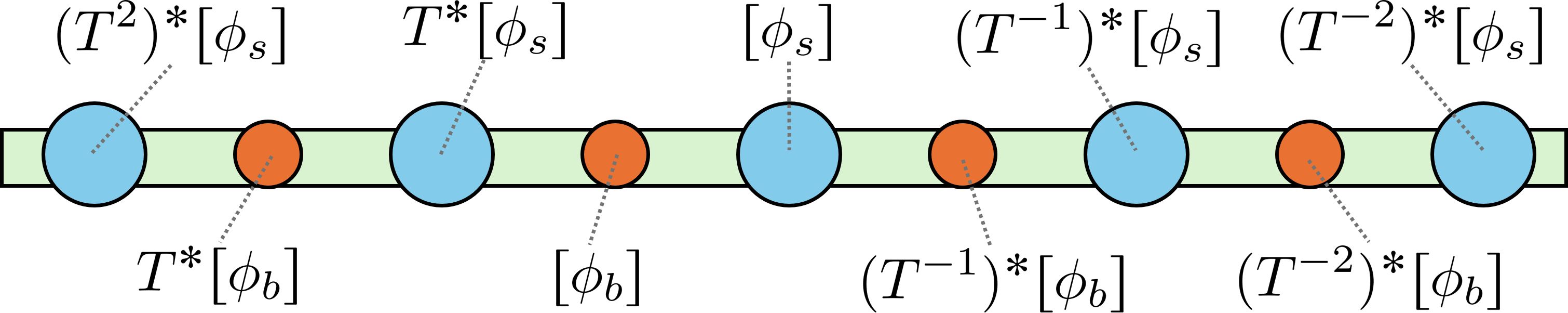}
            \caption{}
            \label{fig:cellulation1DReflection0Cell}
        \end{subfigure}
        \caption{Defect network for a (1+1)D state with translation and reflection symmetry. There are two 1-cells (each is a fundamental domain of the space group) and two 0-cells per unit cell. (a) Assignment of data to the 1-cells (black lines); $[\omega]$ is valued in $\H^2(G_{int},\U1)$. (b) Assignment of data to the 0-cells (blue and orange circles); $[\phi_s]$ is valued in $C^1(G_{int}\times \Z_2^R,\U1)$, while $[\phi_b]$ is valued in $C^1(G_{int}\rtimes \Z_2^R,\U1)$ (see text for further explanation).}
        \label{fig:cellulation1DReflection}
    \end{figure}
    
    We need to change the cellulation of space to be finer than we used in the absence of reflection. Starting from Fig.~\ref{fig:ClusterNetwork}, we divide each unit cell into two 1-cells (joined by a new 0-cell) so that each one is a fundamental domain, and ensure that there is a 0-cell at every nontrivial Wyckoff position (which there is). We then assign data to the 1-cells as shown in Fig.~\ref{fig:cellulation1DReflection}. Note that one of the 0-cells is mapped to itself under $R$, while the other one is left invariant under the composed operation $TR$. Importantly, we are \textit{subdividing} the unit cell compared to Fig.~\ref{fig:ClusterNetwork}, not enlarging the unit cell. This means that translation still permutes $Q_1$ and $Q_2$; enlarging the unit cell would remove this translation action.

    From Fig.~\ref{fig:cellulation1DReflection1Cell}, we see that we must impose
    \begin{equation}
        \omega = \overline{R^\ast\omega}
    \end{equation}
    Once this condition is imposed, it follows that we also need to impose
    \begin{equation}
        \omega = T^\ast \omega
    \end{equation}
    The second condition is unchanged from Sec.~\ref{sec:1DCluster}.

    Since $R$ acts trivially on $G_{int}$, $R^\ast \omega = \omega$. As such, the vanishing of the anomaly requires
    \begin{equation}
        [k] = \overline{R^\ast[k]} = [-k]
    \end{equation}
    This condition happens to be redundant with Eq.~\ref{eqn:ClusterAnomaly}, so reflection does not affect the strong SPT data.

    More interestingly, the weak SPT data is modified significantly. On the $0$-cell which is mapped to itself under reflection, we place an element 
    \begin{equation}
        [\phi_s] \in \H^1(\Z_N \times \Z_N \times \Z_2, \U1) =  \Z_{N}^2 \times \Z_2
    \end{equation}
    where $s$ stands for ``site" and the classification follows from the fact that $\Z_2$ acts \textit{unitarily} on $\U1$ since reflection does not reverse the spacetime orientation of the 0-cell. (We are slightly abusing language - strictly speaking the data on the 0-cells forms an $\H^1(\Z_N \times \Z_N \times \Z_2,\U1)$ torsor, but we can just pick some reference trivialization of the 1-cell data on the 0-cells.) The direct product is there because the reflection which maps the 0-cell to itself does not interchange the copies of $\Z_N$.
    
    On the other hand, the $0$-cell that is not mapped to itself under reflection lives on a bond of the lattice, and it is instead mapped to itself under $TR$, which also generates a $\Z_2$ subgroup of the space group which we denote $\Z_2^{TR}$. There is a key difference, though - since $T$ exchanges the copies of $\Z_N$, so does $TR$. Hence we need to place an element
    \begin{equation}
        [\phi_b] \in \H^1(\Z_N \times \Z_N \rtimes \Z_2, \U1) = \Z_N \times \Z_2
    \end{equation}
    on this 0-cell, where $b$ stands for ``bond," and the semidirect product accounts for the action of $TR$ on the internal symmetries. Again, $TR$ acts unitarily on the 0-cell data.

    In total, then, the weak SPT data is naively given by 
    \begin{equation}
        \H^1(\Z_N \times \Z_N \times \Z_2, \U1) \times \H^1(\Z_N \times \Z_N \rtimes \Z_2, \U1) = \Z_{N}^3 \times \Z_2
    \end{equation}

    \begin{figure}
        \centering
        \includegraphics[width=0.5\linewidth]{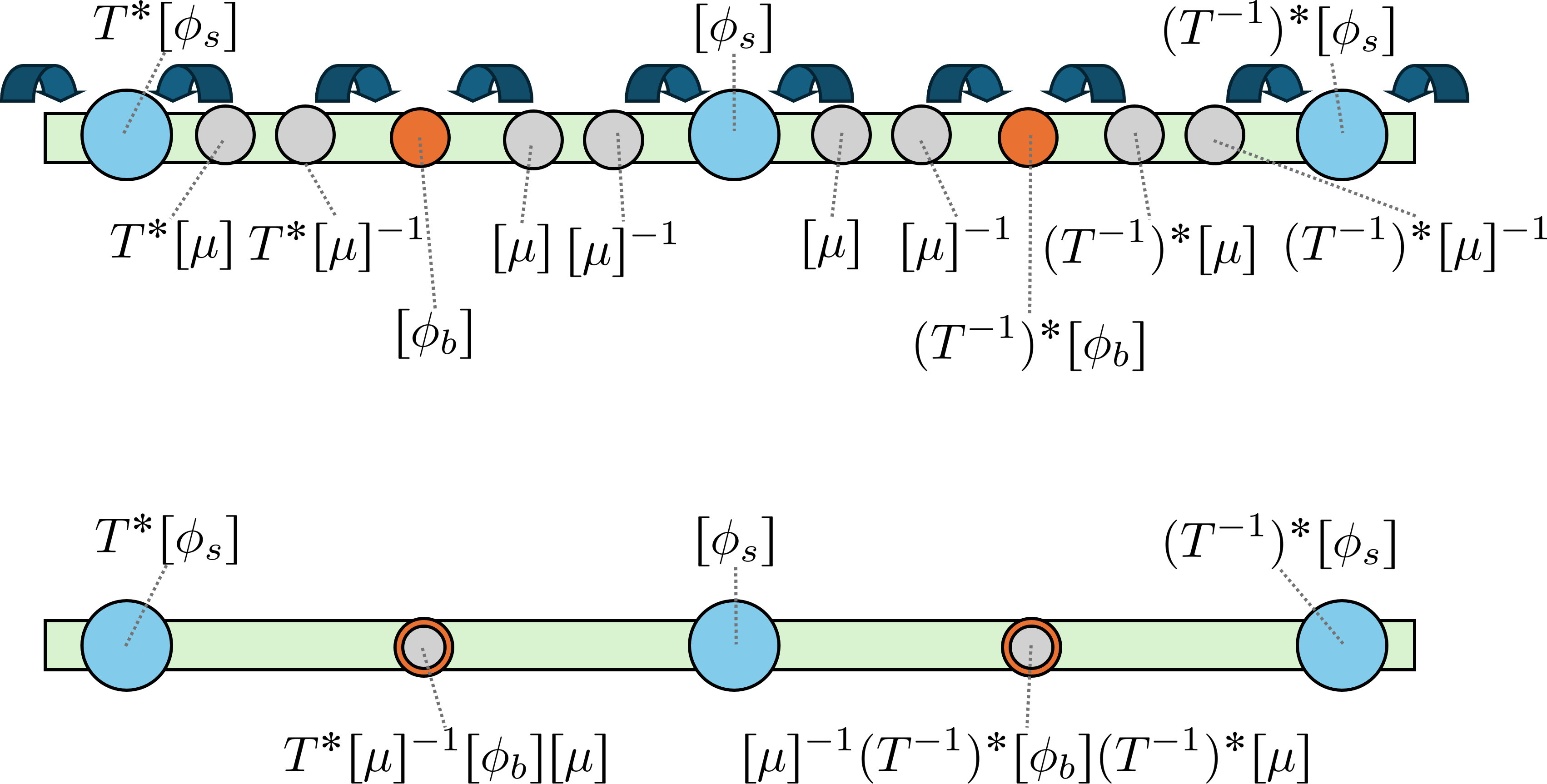}
        \caption{Equivalence of weak SPT data for cluster states with reflection symmetry. First (top) nucleate pairs of (0+1)D SPTs $[\mu]$ and $[\mu]^{-1}$ (grey circles) on a reference 1-cell and use reflection and translation symmetry to place them on all 1-cells (green). Then move them (blue arrows) to the 0-cells (blue and orange circles) to obtain equivalent weak SPT data (bottom).}
        \label{fig:reflectionEquivalence}
    \end{figure}

    We now need to deal with the equivalences like Eq.~\ref{eqn:0CellEquivalence}. The relevant physical process is shown in Fig.~\ref{fig:reflectionEquivalence}. We nucleate a pair of (0+1)D SPTs $[\mu],[\mu]^{-1} \in \H^1(\Z_N \times \Z_N,\U1)$ on a 1-cell, and use spatial symmetries to place appropriate data on all 1-cells. Importantly, reflections act trivially on $\Z_N \times \Z_N$ but antiunitarily on $\U1$ (a reminder that in our convention, we track this by letting $R^\ast$ act trivially and manually accounting for the complex conjugation), so we have to track inverses carefully. We then merge this data into the 0-cells. As can be seen in Fig.~\ref{fig:reflectionEquivalence}, this merges trivial data into $[\phi_s]$. Importantly, $T^\ast[\mu]^{-1}[\mu]$ is invariant under the action of $TR$ and can therefore be viewed as an element of $\H^1(\Z_N \times \Z_N \rtimes \Z_2^{TR},\U1)$. It is therefore well-defined to quotient out the weak SPT data $[\phi_b]$ by the subgroup consisting of all elements $T^\ast[\mu]^{-1}[\mu]$. This is straightforwardly checked to be the whole $\Z_N$ subgroup of $\H^1(\Z_N \times \Z_N \rtimes \Z_2,\U1)$. Hence the actual weak SPT data is given by
    \begin{equation}
        \Z_{N}^2 \times \Z_2.
    \end{equation}

     Before concluding, it is enlightening to consider the action of translation on the weak SPT data. Consider the data $[\phi_s] \in \H^1(\Z_N \times \Z_N \times \Z_2,\U1)$ which lives on the 0-cell on lattice site 0; the same argument will hold for $[\phi_b]$. By translation symmetry, we need to place $T^\ast [\phi_s]$ on site 1. Interestingly, translation does not map $\Z_2^R$ to itself; instead it maps $R \rightarrow T^2R$. Strictly speaking, then, $T^\ast [\phi_s]$ does not live in the same group as $[\phi_s]$; instead, it lives in a copy of $\H^1(\Z_N\times \Z_N \times \Z_2,\U1)$ where the $\Z_2$ is generated by $T^2R$, not by $R$. This makes sense, since $R$ does not map site 1 to itself; the space group element which does map site 1 to itself is $T^2R$. It is straightforward to check that $(T^2R)^2=1$ and that $T^2R$ has the same action on $G_{int}$ as $R$ does; as such, $T^\ast [\phi_s]$ lives in a group that is \textit{isomorphic} to the group that $[\phi_s]$ lives in, which is a nice consistency check. A similar argument holds for translates of $[\phi_b]$.
    
    \subsection{\texorpdfstring{$\U1$}{U(1)} charge and \texorpdfstring{$\Z_L$}{ZL} dipolar symmetries}
        \label{sec:LSM}

        Dipolar symmetries often interact nontrivially with finite system sizes. Our formalism has so far taking the system size to be infinite; we now give an example that shows how our formalism accounts for finite system size.
        
        Ref.~\cite{burnell2024} considers a system of length $L$ with periodic boundary conditions and $\U1$ charge symmetry, but the dipolar symmetry is only $\Z_L$-valued due to the finite system size. They obtain a constraint relating the dipole filling per unit cell to the charge filling per unit cell in a gapped system, namely, that the dipole moment per unit cell $\nu_D$ is constrained by the charge per unit cell $\nu$ via
        \begin{equation}
            \nu_D - \nu \frac{L+1}{2} \in \Z.
        \end{equation}
        We presently reproduce that constraint by classifying SPTs in our language.
        
        The internal symmetry group is $G_{int}=\U1 \times \Z_L$. If $Q$ generates the $\U1$ symmetry and $D$ generates the $\Z_L$ symmetry, then
        \begin{align}
            T_x Q T_x^{-1} &= Q\\
            T_x D T_x^{-1} &= e^{2\pi i Q/L}D
        \end{align}
        We can perform our usual analysis. Using the K\"unneth formula, we find that $\H^2(G_{int},\U1)=0$, so all the 1-cell data is trivial. The 0-cell data is then naively classified by $\H^1(G_{int},\U1) = \Z \times \Z_L$, and in particular $\H^1(G_{int},\U1) = \H^1(\U1,\U1) \times \H^1(\Z_L,\U1)$, that is, we have pure charge weak SPTs and pure dipolar weak SPTs. Applying Eq.~\ref{eqn:0CellEquivalence}, just as before, trivializes the pure dipolar weak SPTs, but this trivialization is irrelevant to the constraint we are aiming for.
        
        To obtain the constraint, we do a similar analysis to Sec.~\ref{sec:1DDipole0Cell}. An element of $\H^1(G_{int},\U1)$ is labeled $(k_Q, k_x)$ where $k_Q \in \Z$ and $k_x \in \Z_L$. Just as in Sec.~\ref{sec:1DDipole0Cell}, the total dipole charge (which is only defined modulo $L$) is given by
        \begin{equation}
            Q_d = \sum_{j=1}^L (k_x + j k_Q) = k_xL + k_Q \frac{L(L+1)}{2} \mod L
        \end{equation}
        Note that this equation is where we use the fact that the system size matches the periodicity of the dipolar symmetry; the sum runs up to the system size (equal to $L$), and the result is taken modulo the periodicity of the dipolar symmetry (also $L$). The analysis before this does not use that fact.
        
        Since $k_x \in \Z_L$, we can drop the $k_xL$ term from this equation (note that this reflects the fact that the pure dipolar weak SPTs are themselves trivial). Interpreting $k_Q$ as the charge filling fraction $\nu$ and defining the dipole filling fraction $\nu_D = Q_D/L$, we immediately obtain the desired constraint
        \begin{equation}
            \nu_d -\nu \frac{L+1}{2} = 0 \mod 1
        \end{equation}
        In our formalism, this constraint does not arise from a 't Hooft anomaly. Instead, it merely arises from inputting the relationship between the system size and the (formally defined) symmetry group.

    \subsection{General \texorpdfstring{$\Z_N$}{ZN} multipolar symmetries in (1+1)D}

        We now classify SPTs protected by $\Z_N$ multipolar symmetries in (1+1)D.

        A multipolar symmetry group in (1+1)D is given by a collection of $\Z_N$-valued polynomials in the spatial coordinate $x$ with degree at most $n$. (Note that all polynomials of degree $\geq N$ can be rewritten as polynomials of degree $<N$, since $x^N=x \mod N$ for any integer $x$. We therefore assume $n<N$.) For example, this occurs for a scalar field which transforms under a symmetry $Q_f$ as $\phi(x) \rightarrow \exp(2\pi if(x)/N)\phi(x)$, where $f(x)$ is any $\Z_N$-valued polynomial of degree at most $n$. Different $n$ produce different multipolar symmetry groups; obviously $n=0$ corresponds to charge conservation, and $n=1$ corresponds to dipole conservation. It is straightforward to show that imposing translation symmetry requires us to consider polynomials of degree at most $n$, not just polynomials of some fixed degree. Imposing multipolar symmetry therefore gives us an internal symmetry group $G_{int}=\Z_N^{n+1}$.
        
        We need to determine the action of translation on $G_{int}$. Although monomials of degree at most $n$ are the most natural generators of the multipolar group, they have messy transformation rules under translation. We instead choose a convenient set of generating polynomials
        \begin{equation}
            f_k(x) = \frac{1}{k!}x(x-1)(x-2)\cdots(x-k+1)
        \end{equation}
        for $0\leq k \leq n$. It is straightforward to check that $f_k$ is $\Z_N$-valued for integer $x$, $f_k$ has degree $k$, and translation acts by
        \begin{equation}
            T(f_k(x))=f_k(x+1)=f_k(x)+f_{k-1}(x)
            ~\label{eqn:translationOnMultipole1D}
        \end{equation}
        for $k>0$, where we have set the lattice constant to 1 for simplicity. (Obviously $T(f_0)=f_0$ since $f_0=1$.) We notate an element $g \in G_{int}$ by $g=(g_0,g_1,\ldots,g_{n-1},g_n)$, where $g_k$ is the (additive) $\Z_N$-valued coefficient of $f_k$. Converting Eq.~\ref{eqn:translationOnMultipole1D} to this notation, we see that
        \begin{equation}
            T(g)= (g_0+g_1,g_1+g_2,\ldots, g_{n-1}+g_n,g_n)
        \end{equation}

        Since $G_{int}$ is Abelian, representative 2-cocycles for the generators of $\H^2(G_{int},\U1)$ all take the form
        \begin{equation}
            \omega_{ij}((g_0,g_1,\ldots,g_n),(h_0,h_1,\ldots,h_n)) = e^{\frac{2\pi i}{N}g_ih_j}
        \end{equation}
        where all choices $i<j$ generate all possible 2-cocycles. The action of translation is
        \begin{align}
            T^\ast \omega_{ij} &= \begin{cases}
                e^{\frac{2\pi i}{N}(g_i+g_{i+1})(h_j+h_{j+1})} & j<n \\
                e^{\frac{2\pi i}{N}(g_i+g_{i+1})h_j} & j=n
            \end{cases} \\
            &=\omega_{ij}\omega_{i+1,j}\omega_{i,j+1}\omega_{i+1,j+1}
        \end{align}
        where in the last line, we define shorthand where $\omega_{ij}=1$ if $j>n$ or $i=j$. This is manifestly translation-invariant if $i=n-1,j=n$, and otherwise clearly not translation-invariant. However, various combinations of the generators are translation-invariant. In particular, let $\ell$ be a positive integer. Then the cocycle
        \begin{equation}
            \omega_\ell = \prod_{i=1}^\ell \prod_{j=1}^i \left(\omega_{n-2\ell+i,n-j+1}\right)^{(-1)^{\ell-j}\binom{\ell-j}{\ell-i}},
        \end{equation}
        where $\binom{\ell-j}{\ell-i}$ is a binomial coefficient, can be tediously but explicitly checked to be translation-invariant, provided that all the $\omega_{ij}$ that appear in the product are well-defined for the $n$ in question. Since $\omega_{ij}$ only exist for $0\leq i<n$ and $0 < j \leq n$, it is straightforward to check that $\omega_\ell$ is well-defined for $\ell \leq (n+1)/2$. Each such $\omega_\ell$ determines a non-anomalous strong MSPT.

        The end result is that we get a strong MSPT classification $\Z_N^{\left\lfloor \frac{n+1}{2}\right\rfloor}$. This classification matches that given in Ref.~\cite{EbisuMultipolar} via matrix product state methods. 
        
        Finally, the weak SPT data is naively given by $\H^1(G_{int},\U1)=\Z_N^{n+1}$. Representative generating cocycles are given by
        \begin{equation}
            \mu_{\lbrace k_m\rbrace}(g) = e^{\frac{2\pi i}{N}\sum_m k_m g_m}
        \end{equation}
        Applying the equivalence Eq.~\ref{eqn:0CellEquivalence}, we straightforwardly find that all of the $k_m$ for $m>0$ label trivial data; only the pure charge, i.e., $k_0$, gives nontrivial data. Hence the weak SPT classification is actually just one copy of $\Z_N$.

    \subsection{Exponentially modulated symmetries}

        Although much of this work focuses on multipolar symmetries, our formalism is not limited to such symmetries; as an example, we can find the SPT classification for the ``exponentially modulated symmetries" considered in Ref.~\cite{han2023modulated}. Other exponentially modulated symmetries have been considered in, for example, Refs.~\cite{SalaModulatedDynamics,LianQuantumBreakdown,LehmannFragmentation,WatanabeModulatedTC}; the formalism extends straightforwardly to those cases as well.
        
        Ref.~\cite{han2023modulated} considers models consisting of a chain of $\Z_N$ degrees of freedom with the following Hamiltonian:
        \begin{equation}
            H = -\sum_{i \text{ odd}}Z^{\dagger}_{i-1}X_iZ_{i+1}^a - \sum_{i \text{ even}} Z_{i-1}^a X_i Z^\dagger + \text{ h.c.}
        \end{equation}
        Note that the unit cell consists of two sites.
        When the system size is infinite, there are two modulated symmetries:
        \begin{align}
            A &= \bigotimes_{j}X_{2j+1}^{a^j}\\
            B &= \bigotimes_j X_{2j}^{a^{-j}}
        \end{align}
        The behavior of the model depends on the number-theoretic relationship between $a$ and $N$.
        
        \subsubsection{\texorpdfstring{$a$}{a} and \texorpdfstring{$N$}{N} are coprime}
        The most interesting behavior happens when $a$ and $N$ are coprime. In this case, the $m$th power of $A$ (resp. $B$) involves operators $X^{ma^j}$ (resp. $X^{ma^{-j}}$) for all $j$, and in particular $j=1$ (resp. -1). It immediately follows that $A$ and $B$ are both $\Z_N$ symmetries, and it is not hard to check that 
        \begin{align}
            T(A) &= A^a\\
            T(B) &= B^{a^{-1}}
            \label{eqn:translationOnExponential}
        \end{align}
        This means that for a symmetry element $(g_A, g_B) \in \Z_N \times \Z_N$ with additive notation,
        \begin{equation}
            T((g_A,g_B)) = (a g_A, a^{-1}g_B)
        \end{equation}
        The 1-cell data for MSPTs in this case are given by elements of $\H^2(\Z_N \times \Z_N,\U1)$, which has the representative cocycles given in Eq.~\ref{eqn:ZnZnCocycle}. These are manifestly invariant under the action of $T^\ast$, so the strong SPT classification is $\Z_N$. There are no further anomalies on the 0-cells, and the 0-cell (weak SPT) data is naively classified by $\H^1(\Z_N \times \Z_N,\U1) = \Z_N \times \Z_N$. Applying the equivalence Eq.~\ref{eqn:0CellEquivalence}, if $(k_A, k_B) \in \H^1(\Z_N \times \Z_N,\U1)$, then we straightforwardly find
        \begin{equation}
            \left((a-1)k_A,(a^{-1}-1)k_B\right) \sim (0,0)
        \end{equation}
        Quotienting out by this subgroup, we obtain weak SPT data
        \begin{equation}
            \Z_{(a-1,N)} \times \Z_{(a^{-1}-1,N)}
        \end{equation}
        where $(x,y)$ means the greatest common divisor of $x$ and $y$.

        \subsubsection{Other cases}

        Next consider $a = 0 \mod \text{rad}(N)$, where rad$(N)$ is the product of distinct prime factors of $N$. Then $a$ does not have a multiplicative inverse modulo $N$, and one can check that every positive power of $A$ (resp. $B$) is only well-defined if the system has a boundary on the left (resp. right), i.e., if $j$ is always nonnegative (resp. nonpositive), and furthermore are each only supported on finite piece of the system. This is basically a 0+1D SPT since the symmetry only acts on a finite number of degrees of freedom, and the boundary breaks translation symmetry, so this case is not of interest to us as a MSPT. We will not consider it further.
        
        Finally, consider the case where $a$ and $N$ are not coprime, but also $a \neq 0 \mod \text{rad}(N)$. Then as discussed in Ref.~\cite{han2023modulated}, we can define $N_a$ to be the largest divisor of $N$ which is coprime to $a$. The only symmetries that are well-defined are $\tilde{A} = A^{N/N_a}$ and $\tilde{B} = B^{N/N_a}$, which form a $\Z_{N_a} \times \Z_{N_a}$ modulated symmetry group.  It is straightforward to check that $\tilde{A}$ and $\tilde{B}$ obey Eq.~\ref{eqn:translationOnExponential}, so we just reduce to the case where $a$ and $N$ are coprime, but with $N$ replaced by $N_a$. Hence the strong SPT classification is $\Z_{N_a}$ and the weak classification is $\Z_{N_a} \times \Z_{N_a}$.

\section{Classification of dipolar SPTs on the square lattice}
    \label{sec:2DDipolar}

    We now classify SPTs protected by dipolar symmetries with square lattice translation symmetry. We always include translation symmetry $T_x$ and $T_y$, along with charge conservation symmetry $Q$ in the full symmetry group. However, we can now consider cases where the $x$ component of dipole is conserved but not the $y$ component, or where both components are conserved. (We could also consider a single generic direction for a conserved component of the dipole moment, but our results here illustrate the principle.)

    The classification is of interest in itself, but we highlight that nontrivial anomalies can appear on the 0-cells when both $x$ and $y$ dipole symmetries are included, even in the absence of point group symmetry. This feature does not occur for unmodulated symmetries.

    \subsection{\texorpdfstring{$\Z_N$}{ZN} dipolar symmetry in one direction}

    Without loss of generality we assume that the $x$ component of dipole moment is conserved. Hence the internal symmetry group is $G_{int} = \Z_N^{(Q)}\times \Z_N^{(x)}$ (where the $x$ represents the $x$ component of dipole moment). Translation acts in the usual way
    \begin{align}
        T_{x,y}QT_{x,y}^{-1} &= Q\\
        T_xD_xT_{x}^{-1} &= QD_x\\
        T_yD_xT_y^{-1} &= D_x
    \end{align}
    Letting $g = (g_Q, g_x) \in \Z_N \times \Z_N$, where we are anticipating later notation, the action of translation sends $T_x((g_Q,g_x))= ([g_Q+g_x],g_x)$ in this notation. 
    
    \subsubsection{2-cell data}
    We need to choose $\omega \in \H^3(\Z_N \times \Z_N,\U1) = \Z_N^3$. Representative cocycles for the generators of $\Z_N^3$ are:
    \begin{align}
        \omega_I^{(Q)} &= e^{\frac{2\pi i}{N^2}g_Q\left(h_Q+k_Q-[h_Q+k_Q]\right)}\\
        \omega_I^{(x)} &= e^{\frac{2\pi i}{N^2}g_x\left(h_x+k_x-[h_x+k_x]\right)}\\
        \omega_{II} &= e^{\frac{2\pi i}{N^2}g_Q\left(h_x+k_x-[h_x+k_x]\right)}
    \end{align}
    where $[g+h]$ denotes addition mod $N$ and the subscript denotes the ``type" of the cocycle. Obviously all of these are invariant under $T_y^\ast$, so we just need to focus on $T_x^\ast$.
    
     Clearly $T^\ast\omega_I^{(x)}=\omega_I^{(x)}$, so this cocycle data is non-anomalous. For $\omega_{II}$, we compute that
    \begin{equation}
        T_x^\ast \omega_{II} = e^{\frac{2\pi i}{N^2}\left([g_Q+g_x]\right)\left(h_x+k_x-[h_x+k_x]\right)}
    \end{equation}
    Using the fact that $h_x+k_x-[h_x+k_x]$ is always equal to 0 or $N$, one can check that it is safe drop the brackets on $[g_Q+g_x]$. Therefore,
    \begin{equation}
        T_x^\ast\omega_{II} = \omega_I^{(x)}\omega_{II}
    \end{equation}
    and therefore $T_x^\ast\omega_{II}$ is not in the same cohomology class as $\omega_{II}$.
    
    Finally, we can check $\omega_{I}^{(Q)}$. To this end, we need to determine the cohomology class of
    \begin{equation}
        T_x^\ast \omega_I^{(Q)} = e^{\frac{2\pi i}{N^2}(g_Q+g_x)\left([h_Q+h_x]+[k_Q+k_x]-[h_Q+h_x+k_Q+k_x]\right)}
    \end{equation}
    We use the coboundary generated by
    \begin{equation}
        \nu(g,h) = e^{\frac{2\pi i}{N^2}\left((g_Q+g_x)(h_Q+h_x-[h_Q+h_x])+g_Qh_x-(g_Q+h_Q-[g_Q+h_Q])(g_x+h_x)\right)} \label{eqn:dnuOmegaII}
    \end{equation}
    to see that
    \begin{equation}
        T_x^\ast \omega_I^{(Q)}d\nu = \omega_I^{(Q)}\omega_I^{(x)}\omega_{II}^2
    \end{equation}
    Hence we conclude that $\omega_I^{(Q)}$ is not invariant under $T_x^\ast$ and is therefore anomalous data for the defect network.
    
    To summarize, nontrivial anomalies are generated by
    \begin{align}
        T_x^\ast \left[\omega_I^{(Q)}\right]&= \left[\omega_I^{(Q)}\right]\left[\omega_I^{(x)}\right]\left[\omega_{II}\right]^2 \\
        T_x^\ast \left[\omega_{II}\right] &= \left[\omega_{II}\right]\left[\omega_I^{(x)}\right]
    \end{align}
    
    By inspection, the choices of 2-cell data $[\omega]$ which do not create an anomaly at the 1-cell are $[\omega_I^{(x)}]$ for any $N$ and $\left(\left[\omega_I^{(Q)}\right]\left[\omega_{II}\right]\right)^{N/2}$. for $N$ even. Hence the strong SPT classification is $\Z_N \times \Z_{(2,N)}$, where as usual $(2,N)$ represents the greatest common divisor of $2$ and $N$.
    
    \subsubsection{1-cell data}    
    The symmetry that preserves each 1-cell is just $G_{int}$. If we choose the 2-cell data $\omega = \omega_I^{(x)}$, then $\omega$ is manifestly invariant on the nose under both $T_x^\ast$ and $T_y^\ast$, so data on each 1-cell is labeled by an actual element of $\H^2(G_{int},\U1)$. That is, we dress the 1-cells with a (1+1)D dipolar SPT. As in Sec.~\ref{sec:LowerCell}, we have independent choices for the horizontal and vertical 1-cells.
    
    For even $N$, we could also choose $[\omega]= \left(\left[\omega_I^{(Q)}\right]\left[\omega_{II}\right]\right)^{N/2}$. In this case, one can use the above results to show that $T_x^\ast \omega/\omega$ is trivialized by $\nu^{N/2}$, where $\nu$ is given in Eq.~\ref{eqn:dnuOmegaII}, while $T_y^\ast \omega = \omega$. Obviously shifting $\nu$ by a 2-cocycle will not change the fact that $\nu$ trivializes $T_x^\ast \omega/\omega$.

    Hence, the naive 1-cell data is given by
    \begin{align}
        \mu_h &= z_h \nonumber\\
        \mu_v &= \begin{cases}
                \nu z_v & N \text{ even and } [\omega]= \left(\left[\omega_I^{(Q)}\right]\left[\omega_{II}\right]\right)^{N/2}\\
                z_v & \text{else}
        \end{cases}
        \label{eqn:QDx1CellData}
    \end{align}
    where $z_h$ and $z_v$ are 2-cocycles. The data $\mu_h$ is classified by $\H^2(\Z_N,\U1)=\Z_N$, while the data $\mu_v$ forms an $\H^2(\Z_N,\U1)=\Z_N$ torsor. 

    We need to deal with any equivalences from Eq.~\ref{eqn:1CellEquivalence}.  We can reuse the (1+1)D classification results in Sec.~\ref{sec:1DZnDipolar}, to check that 2-cocycles $z_v$ and $z_h$ that appear in Eq.~\ref{eqn:QDx1CellData} are invariant (as cohomology classes) under $T_x^\ast$ (and are obviously invariant under $T_y^\ast$, which acts trivially). Plugging into Eq.~\ref{eqn:1CellEquivalence}, we find no nontrivial equivalences.
    
    To summarize, provided we check for anomalies on the 0-cells, the 1-cells contribute a $\Z_N \times \Z_N$ weak SPT classification.
    
    \subsubsection{0-cell data}
    
    There is no extra symmetry at the 0-cell. Given that $T_y^\ast \nu = \nu$ on the nose, we can insert the data in Eq.~\ref{eqn:QDx1CellData} into Eq.~\ref{eqn:squareLattice0CellAnomaly} and immediately see that there is no anomaly on the 0-cells for any non-anomalous choice of $\omega$.
    
    Distinct 0-cell data then naively forms a $\H^1(G_{int},\U1)=\Z_N \times \Z_N$ torsor. These are analogous to the 1+1D 0-cell data, namely, they tell us about charge and neutral dipole per unit cell. By exactly the same argument as in 1+1D, we need to quotient this by the relations
    \begin{equation}
        T_x^\ast[\phi] \sim T_y^\ast[\phi] \sim [\phi]
    \end{equation}
    This removes the pure dipolar copy of $\Z_N$ weak SPT data.
    
    To conclude, with translation symmetry SPTs with one direction of dipole symmetry are classified by $\Z_N^4 \times \Z_{(2,N)}$. Two of those copies of $\Z_N$ are 1D weak dipole SPTs, one copy is a 0D weak SPT, and one copy is a strong (2+1)D dipole SPT. The $\Z_{(2,N)}$ is an extra strong index.

    \subsection{\texorpdfstring{$\U1$}{U(1)} dipolar symmetry in one direction}

    A straightforward application of the K\"unneth formula tells us that
    \begin{equation}
        \H^3(\U1^2,\U1) = \Z^3
    \end{equation}
    The explicit cocycles generalize the discrete case; with $i=Q,x$ representing charge and dipolar symmetries, we have:
    \begin{align}
        \omega_I^{(i)}(g,h,k) &= e^{\frac{i}{2\pi}\theta_{g,i}(\theta_{h,i}+\theta_{k,i}-[\theta_{h,i}+\theta_{k,i}])}\\
        \omega_{II}(g,h,k) &= e^{\frac{i}{2\pi}\theta_{g,Q}(\theta_{h,x}+\theta_{k,x}-[\theta_{h,x}+\theta_{k,x}])}
    \end{align}
    where $g_i = e^{i\theta_i}$ and the brackets indicate addition modulo 2$\pi$.
    
    The computation of how translations act on these cocycles is the same as in the $\Z_N$ case if we make the replacement $g_i = 2\pi \theta_{g,i}/N$ (implicitly there is an $N \rightarrow \infty$ limit, but the $N$s all cancel after this replacement). Hence $\omega_{II}^{(x)}$ generates a $\Z$ classification of strong (2+1)D $\U1$ dipolar SPTs. Importantly, no power of $\left[\omega_I^{(Q)}\right]\left[\omega_{II}^{(x)}\right]$ is invariant under $T_x^\ast$, so there is no extra factor of $\Z_2$ like there is for finite even $N$.
    
    Since $\omega_{II}^{(x)}$ is invariant on the nose under $T_x^\ast$ and $T_y^\ast$, distinct 1-cell data is classified by elements of $\H^2(\U1^2,\U1)$, which is trivial. Hence there is a unique, non-anomalous choice for the 1-cell data, for which the trivial cocycle is a convenient representative.
    
    With the aforementioned choice of representative for the 1-cell data, which is obviously invariant under translations, the 0-cell data is naively classified by elements of $\H^1(\U1 \times \U1,\U1) = \Z \times \Z$, which corresponds to a weak SPT with a charge or a dipole per unit cell. As in the discrete case, the dipole per unit cell is equivalent to the trivial phase, dropping the classification down to $\Z$.
    
    To conclude, $\U1$ strong dipolar SPTs with one direction of dipolar symmetry have a $\Z$ classification, and the corresponding weak SPTs have a $\Z$ classification.
    
    By comparison, an unmodulated $\U1^2$ symmetry would have a $\Z^3$ classification for strong SPTs and a $\Z^2$ classification for weak SPTs.

    \subsection{\texorpdfstring{$\Z_N$}{ZN} dipolar symmetry in two directions}
    \label{sec:ZNDxDy}

    Next we can impose $\Z_N^3$ symmetry, generated by $Q,D_x,D_y$. This example is particularly interesting because anomalies appear on the 0-cells despite the fact that there is no point group symmetry, which does not happen in the unmodulated case.
    
    In additive $\Z_N$ notation,
    \begin{align}
        T_x((g_Q,g_x,g_y))&=([g_Q+g_x],g_x,g_y) \nonumber \\
        T_y((g_Q,g_x,g_y))&=([g_Q+g_y],g_x,g_y)
        \label{eqn:2DDipoleTranslation}
    \end{align}
    
    \subsubsection{Strong SPT (2-cell) data}
        We need to choose an element $[\omega] \in \H^3(\Z_N^3,\U1)=\Z_N^7$. The representative cocycles are, for $i=Q,x,y$ representing $Q$, $D_x$, and $D_y$ respectively,
        \begin{align}
                \omega_I^{(i)} &= e^{\frac{2\pi i}{N^2}g_i\left(h_i+k_i-[h_i+k_i]\right)}\\
            \omega_{II}^{(ij)} &= e^{\frac{2\pi i}{N^2}g_i\left(h_j+k_j-[h_j+k_j]\right)}\\
            \omega_{III} &= e^{\frac{2\pi i}{N}g_Qh_xk_y}
        \end{align}
        where for $\omega_{II}$, $i<j$ generates all cocycles.
        
        We can repeat the results of the previous section to see that $\omega_I^{(x)}$ and $\omega_I^{(y)}$ are non-anomalous and $\omega_{I}^{(Q)}$, $\omega_{II}^{(Qx)}$, and $\omega_{II}^{(Qy)}$ are all anomalous. From Eq.~\ref{eqn:2DDipoleTranslation}, we immediately see that $\omega_{II}^{(xy)}$ is non-anomalous. This leaves $\omega_{III}$ to be checked. It is immediate to see that
        \begin{align}
            T_x^\ast\omega_{III}/\omega_{III} &= e^{\frac{2\pi i}{N}g_xh_xk_y}\\
            T_y^\ast\omega_{III}/\omega_{III} &= e^{\frac{2\pi i}{N}g_yh_xk_y}.
        \end{align}
        A careful computation shows that
        \begin{align}
            T_x^\ast \omega_{III}/\omega_{III} &= \begin{cases}
                d\left(e^{\frac{2\pi i}{N}\frac{N+1}{2}\left(-g_x^2h_y+g_xh_y\right)}\right) & N \text{ odd}\\
                \left(\omega_{II}^{(xy)}\right)^{N/2} d\left(e^{\frac{2\pi i}{N}\frac{1}{2}\left(g_x^2h_y\right)}\right) & N \text{ even}
            \end{cases}\label{eqn:omegaIIItrivialization}\\
            T_y^\ast \omega_{III}/\omega_{III} &= \begin{cases}
                d\left(e^{\frac{2\pi i}{N}\left(g_x g_y h_y + \frac{N+1}{2}g_x h_y^2\right)}\right) & N \text{ odd}\\
                \left(\omega_{II}^{(xy)}\right)^{N/2} d\left(e^{\frac{2\pi i}{N}\left(g_x g_y h_y + \frac{1}{2}g_x h_y^2 -\frac{1}{2}g_x h_y\right)}\right) & N \text{ even}
            \end{cases}
        \end{align}
        
        For $N$ odd, then, $\omega_{III}$ is non-anomalous, while odd powers of $\omega_{III}$ are anomalous for $N$ even.
    
        To summarize the results of this section, the nontrivial anomalies arise from
    \begin{align}
        T_x^\ast \left[\omega_I^{(Q)}\right] &= \left[\omega_I^{(Q)}\right]\left[\omega_I^{(x)}\right] \left[\omega_{II}^{(Qx)}\right]^2\\
        T_y^\ast \left[\omega_I^{(Q)}\right] &= \left[\omega_I^{(Q)}\right]\left[\omega_I^{(y)}\right] \left[\omega_{II}^{(Qy)}\right]^2\\
        T_x^\ast \left[\omega_{II}^{(Qx)}\right] &= \left[\omega_{II}^{(Qx)}\right]\left[\omega_I^{(x)}\right]\\
        T_y^\ast \left[\omega_{II}^{(Qx)}\right] &= \left[\omega_{II}^{(Qx)}\right]\left[\omega_{II}^{(xy)}\right]\\
        T_x^\ast \left[\omega_{II}^{(Qy)}\right] &= \left[\omega_{II}^{(Qy)}\right]\left[\omega_{II}^{(xy)}\right]\\
        T_y^\ast \left[\omega_{II}^{(Qy)}\right] &= \left[\omega_{II}^{(Qy)}\right]\left[\omega_I^{(y)}\right]\\
        T_x^\ast \left[\omega_{III}\right] =  T_y^\ast \left[\omega_{III}\right] &= \left[\omega_{III}\right]\left[\omega_{II}^{(xy)}\right]^{N/2} \text{ (only for } N \text{ even)}
        \end{align}
        
        For $N$ odd, the non-anomalous choices of $\omega$ consist of any product of $\left[\omega_I^{(x)}\right],\left[\omega_I^{(y)}\right]$, $\left[\omega_{II}^{(xy)}\right]$, and $\left[\omega_{III}\right]$, which forms the group $\Z_N^4$. For $N$ even, the non-anomalous choices of $\omega$ consist of any product of $\left[\omega_I^{(x)}\right],\left[\omega_I^{(y)}\right]$, $\left[\omega_{II}^{(xy)}\right]$, and $\left[\omega_I^{(Q)}\omega_{II}^{(Qx)}\omega_{II}^{(Qy)}\right]^{N/2}\left[\omega_{III}\right]$, which also forms the group $\Z_N^4$.
        
        In either case, the strong SPT classification is $\Z_N^4$. We will show shortly that all of these strong SPTs admit valid weak SPT data, although there are nontrivial constraints on that weak SPT data.
    
    \subsubsection{1-cell data}
    
        As in the previous case, the symmetry that preserves the 1-cell is still $G_{int}=\Z_N^3$.
        
        Let us characterize the data on the 1-cells for different choices of $\omega$. Note that we have independent choices of data on the horizontal and vertical 1-cells.
    
        \underline{Case 1}: If the 2-cell data is $\omega = \left[\omega_I^{(x)}\right],\left[\omega_I^{(y)}\right],$ or $\left[\omega_{II}^{(xy)}\right]$, then the representative cocycles are invariant on the nose under translations. Hence each of the pieces of 1-cell data $\mu_h$, $\mu_v$ are exactly an element of $\H^2(G_{int},\U1) = \Z_N^3$, and we have a naive $\Z_N^6$ classification of the 1-cell data. We will show shortly that one copy of $\Z_N$ is actually anomalous and two are trivial via Eq.~\ref{eqn:1CellEquivalence}, so the classification is actually $\Z_N^3$.
    
        \underline{Case 2}: $N$ odd, and $\omega = \omega_{III}$. In this case, we need the 1-cell data to obey the nontrivial constraints
        \begin{align}
            \frac{T_x^\ast \omega_{III}}{\omega_{III}} &= d\mu_v\\
            \frac{T_y^\ast \omega_{III}}{\omega_{III}} &= d\mu_h
        \end{align}
        which, as demonstrated in Eq.~\ref{eqn:omegaIIItrivialization}, are solved by
        \begin{align}
            \mu_v &= e^{\frac{2\pi i}{N}\frac{N+1}{2}(-g_x^2h_y+g_xh_y)} z_v(g,h) \nonumber \\
            \mu_h &= e^{\frac{2\pi i}{N}(g_xg_yh_y+\frac{N+1}{2}g_xh_y^2)}z_h(g,h) \label{eqn:fullDipoleOddN1Cell}
        \end{align}
        where $z_{v,h}$ are any elements of $\H^2(G_{int},\U1)=\Z_N^3$. Hence the data naively forms a $\left(\H^2(\Z_N^3,\U1)\right)^2=\Z_N^6$ torsor. As in case 1, one copy will turn out anomalous, and two will be trivial.
        
        \underline{Case 3}: $N$ even, and $\omega = \left(\omega_I^{(Q)}\omega_{II}^{(Qx)}\omega_{II}^{(Qy)}\right)^{N/2}\omega_{III}$. Combining a number of previous results we can find that
        \begin{align}
            \mu_v &= e^{\frac{\pi i}{N}\left(\left(g_Q + g_x\right)\left(h_Q+h_x-[h_Q+h_x]\right)+g_Qh_x-\left(g_Q+h_Q-[g_Q+h_Q]\right)(g_x+h_x)+g_x^2h_y\right)}z_v(g,h)\\
            \mu_h &= e^{\frac{\pi i}{N}\left(\left(g_Q + g_y\right)\left(h_Q+h_y-[h_Q+h_y]\right)+g_Qh_y-\left(g_Q+h_Q-[g_Q+h_Q]\right)(g_y+h_y)\right)}\times \nonumber \\
            & \hspace{1cm}\times e^{\frac{\pi i}{N}\left(2g_xg_yh_y+g_xh_y^2\right)}z_h(g,h)
        \end{align}
        where $z_{v,h}$ are any elements of $\H^2(G_{int},\U1)=\Z_N^3$. Again the data naively forms a $\Z_N^6$ torsor, with the same reductions as in the previous cases.
        
        We finally check for equivalences in all three cases. Using the K\"unneth decomposition, each 1-cell data naively forms a $\H^2(\Z_N^3,\U1)=\Z_N^3$ torsor, with representative 2-cocycles
        \begin{equation}
            \mu^{(ij)} = e^{\frac{2\pi i}{N}g_i h_j}
        \end{equation}
        for $i,j$ ranging over $Q,x,y$ (with $\mu^{(ij)}$ cohomologous to $\mu^{(ji)}$). By similar calculations to the (1+1)D case, the nontrivial transformations are
        \begin{align}
            T_x^\ast \left[\mu^{(Qy)}\right] &= \left[\mu^{(Qy)}\right]\left[\mu^{(xy)}\right]\\
            T_y^\ast \left[\mu^{(Qx)}\right] &= \left[\mu^{(Qx)}\right]\left[\mu^{(xy)}\right]
        \end{align}
        
        Applying Eq.~\ref{eqn:1CellEquivalence}, we see that setting either $[\mu_h]$ or $[\mu_v]$, independently, to $[\mu^{(xy)}]$ is trivial, so only $\Z_N^4$ 1-cell data remains after quotienting this data out. We will now check for the promised anomaly on the 0-cell.
        
        \subsubsection{0-cell data}
        
        We now need to check for any anomalies on the 0-cells. We follow the same case analysis as we did for the 1-cells.
        
        \underline{Case 1}: 
        The 1-cell data is actually classified by the aforementioned cocycles rather than just forming a torsor.
        
        Examining Eq.~\ref{eqn:squareLattice0CellAnomaly}, we see that if $\mu_v=\mu^{(Qx)}$, there is an anomaly unless $\mu_h = \mu^{(Qy)}$, and vice-versa. This removes one copy of $\Z_N$ from the possible weak SPT data; the non-anomalous 1-cell data has a $\Z_N^3$ classification. In these non-anomalous cases, the 0-cell data naively forms a $\H^1(\Z_N^3,\U1)=\Z_N^3$ torsor, which physically, this corresponds to adding in a charge, neutral $x$ dipole, or neutral $y$ dipole per unit cell. As in the previous cases, the pure dipolar 0-cell data is equivalent to the trivial state via Eq.~\ref{eqn:0CellEquivalence}, so the 0-cell data is only a $\Z_N$ torsor.
        
        Note also that $\mu^{(xy)}$ is, as a cocycle, invariant under translations, so the 0-cell data is actually classified by $\Z$ in this case, rather than just being a torsor.
        
        \underline{Case 2}: We observe from Eq.~\ref{eqn:fullDipoleOddN1Cell} that the translation properties of $\mu_{v,h}$ depend only on the 2-cocycles $z_{v,h}$, as the other factors in Eq.~\ref{eqn:fullDipoleOddN1Cell} are translation-invariant on the nose. Hence only the 2-cocycles $z$ contribute to Eq.~\ref{eqn:squareLattice0CellAnomaly}. This reduces us to Case 1, meaning that the 1-cell data has a $\Z_N^3$ classification. The 0-cell data forms a $\Z_N$ torsor as before, corresponding to charge per unit cell.
        
        \underline{Case 3}: A careful calculation shows that, if we ignore the contribution from $z_v$ and $z_h$,
        \begin{equation}
            \mu_v \overline{T_y^\ast \mu_v} T_x^\ast \mu_h \overline{\mu_h} = d\lambda
        \end{equation}
        where
        \begin{equation}
            \lambda = e^{\frac{\pi i}{N}\left([g_Q+g_x]g_y-[g_Q+g_y]g_x+g_Q(g_x-g_y)-g_x g_y\right)} z_0(g)
        \end{equation}
        with $z_0(g)$ any 1-cocycle. Hence there is no anomaly arising from the non-cocycle piece of $\mu$. The non-anomalous 1-cell data forms a $\Z_N^3$ torsor, and the 0-cell data forms a $\Z_N$ torsor as in the previous cases.

        To summarize, the overall classification of SPTs with full $\Z_N$ dipolar symmetry on the square lattice consists of a $\Z_N^4$ strong index and a $\Z_N^4$ weak index.

\subsection{\texorpdfstring{$\U1$}{U(1)} dipolar symmetry in two directions}

    This time the K\"unneth formula tells us
    \begin{equation}
        \H^3(\U1^3,\U1)=\Z^6
    \end{equation}
    In particular, the type-III cocycle present in the discrete case does not exist for the $\U1$ case. It is straightforward to check that the rest of the anomaly calculations are identical in the $\Z_N$ and $\U1$ cases.
    Hence the non-anomalous choices are generated by $\left[\omega_I^{(x)}\right]$, $\left[\omega_I^{(y)}\right]$, and $\left[\omega_{II}^{(xy)}\right]$, forming a $\Z^3$ classification.
    
    Each of our representative 3-cocycles in the non-anomalous classes is invariant on the nose, so 1-cell data is given by an element of $\H^2(\U1^3,\U1)=0$. There are no weak SPTs on the 1-cells with this symmetry.
    
    Finally, choosing the trivial 2-cocycle to represent the trivial data on the 1-cells, we find that the 0-cell data is given by elements of $\H^1(\U1^3,\U1)=\Z^3$. Unsurprisingly, this corresponds to placing a charge, neutral $x$ dipole, or neutral $y$ dipole per unit cell, and as usual Eq.~\ref{eqn:0CellEquivalence} tells us that only the charge is nontrivial. Hence the weak SPT data has a $\Z$ classification.

\section{Conclusions}
\label{sec:conclusions}

We have demonstrated how the defect network formalism allows us to use relatively straightforward calculations in group cohomology to classify MSPTs and to diagnose 't Hooft anomalies. Several natural extensions of our work contribute to a program whose ultimate goal is to gain a full understanding of the role of spatial symmetries in fracton orders. Here we have primarily studied invertible phases where the symmetry group is independent of the system size. To study fractons, we will need to study non-invertible phases and also allow the symmetry group to depend on system size. 

To this end, it is natural to incorporate non-invertible phases into our framework, as defect networks are known to describe symmetry-enriched topological phases (SETs) with unmodulated symmetries~\cite{ElseThorngrenDefectNetworks}. One could study examples where spatial symmetries permute anyons (see, e.g.,~\cite{WenPlaquette,WatanabeModulatedTC,PacePositionDependent,KimUVIR}), or study the interplay of modulated or subsystem symmetries with topological order\cite{YoshitomeModulated}. 

Fractons are connected to subsystem symmetries, where not only does the does the symmetry group change with system size, but the number of \textit{generators} of the symmetry group grows with system size. It would be an important next step to generalize our framework to such cases. In Sec.~\ref{sec:LSM}, we made some progress in this direction by allowing the symmetry group does depend on the system size, although the number of generators does not change; the generalization was straightforward, although the way that the system size appeared was somewhat subtle.

It would also be interesting to carefully elucidate the mathematical structure behind our defect network construction. In the case of unmodulated symmetries, defect networks are tightly related to spectral sequences in equivariant (co)homology~\cite{ElseThorngrenDefectNetworks,ShiozakiSpectralSequence,HuangBlockStates,SongBlockState}; presumably our construction has a similar description, but for slightly different symmetry groups. In particular, a defect network realizes a $G$-equivariant generalized homology class, where $G$ is the full symmetry group (including spatial symmetries).

\acknowledgements
DB acknowledges helpful discussions with Fiona Burnell, Hiromi Ebisu, Dominic Else, Sheng-Jie Huang, Ryohei Kobayashi, Naren Manjunath, Sal Pace, Abhinav Prem, and Takuma Saito, and discussions on a related project with Jintae Kim, Yun-Tak Oh, and Jung Hoon Han. This research was supported in part by the US Office of Naval Research Document Number N0001425GI01144. This research was supported in part by grant NSF PHY-2309135 to the Kavli Institute for Theoretical Physics (KITP). This research was performed in part during a visit to the Okinawa Institute of Science and Technology under the Theoretical Sciences Visiting Program. The views expressed in this article are those of the author and do not reflect the official policy or position of the U.S. Naval Academy, Department of the Navy, the Department of Defense, or the U.S. Government.

\appendix

\section{Exactly solvable defect network model for (1+1)D cluster state}
\label{app:ExactlySolvableClusterDefectNetwork}

In this appendix, we construct an exactly solvable defect network model for the (1+1)D $\Z_2 \times \Z_2$ cluster state and use it to demonstrate how defect network data behaves, particularly to illustrate the equivalence relation Eq.~\ref{eqn:0CellEquivalence}. While there is, of course, a standard Hamiltonian Eq.~\ref{eqn:clusterH2} which implements this cluster state with modulated symmetry, it is illustrative to see the implementation as a defect network explicitly.

On each 1-cell, we place the model~\ref{eqn:clusterH1} with $h=0$ to preserve exact solubility. We choose a finite chain of length $\ell$ for each 1-cell, and simply discard any terms which would have support outside the chain.

\begin{figure}
    \centering
    \includegraphics[width=0.5\linewidth]{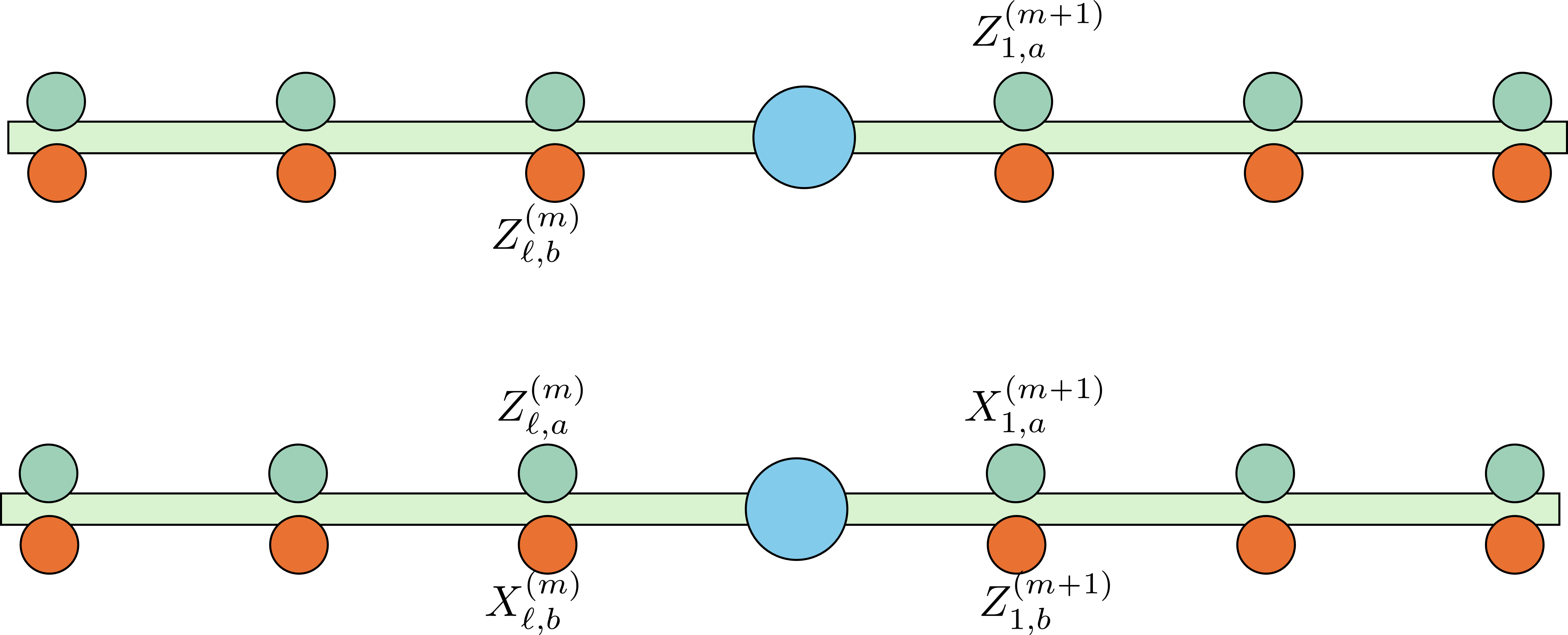}
    \caption{The two terms in the coupling Hamiltonian Eq.~\ref{eqn:exactlySolubleCouple} for the explicit defect network model. The 0-cell (blue) contains no degrees of freedom but separates two neighboring 1-cells (green). On each 1-cell there are two spin-1/2s per microscopic unit cell (teal and orange), and these terms couple neighboring 1-cells.}
    \label{fig:explicitNetworkCouplingH}
\end{figure}

Each 1-cell has two symmetries $Q_a^{(m)}$ and $Q_b^{(m)}$, as written in Eq.~\ref{eqn:clusterSymmH1}, where $m$ labels the 1-cell in question. We now couple two neighboring 1-cells $m$ and $m+1$ with the following terms:
\begin{equation}
    H_{couple} = -Z^{(m)}_{\ell,b}Z^{(m+1)}_{1,a} - Z^{(m)}_{\ell,a}X^{(m)}_{\ell,b}X^{(m+1)}_{1,a}Z^{(m+1)}_{1,b} + \text{ h.c.}
    \label{eqn:exactlySolubleCouple}
\end{equation}
These terms are shown diagramatically in Fig.~\ref{fig:explicitNetworkCouplingH}. It is straightforward to check that this coupling term commutes with every term in the Hamiltonian~\ref{eqn:clusterH1} when $h=0$ and that the  two terms in Eq.~\ref{eqn:exactlySolubleCouple} commute.

Clearly the first term in Eq.~\ref{eqn:exactlySolubleCouple} breaks $Q_b^{(m)}$ and $Q_a^{(m+1)}$ but preserves the product, while the second term breaks $Q_a^{(m)}$ and $Q_b^{(m+1)}$ but preserves the product. This is exactly the modulated symmetry breaking in Eq.~\ref{eqn:modulatedCoupling}. We see that in the presence of $H_{couple}$, the remaining symmetry $G_m$ is
\begin{align}
    Q_1 &= \prod_{m \text{odd}}Q_a^{(m)}Q_b^{(m+1)}\\
    Q_2 &= \prod_{m \text{even}}Q_a^{(m)}Q_b^{(m+1)}
\end{align}
and translation $T$ by $\ell$ microscopic sites. Clearly $T(Q_a^{(m)} = Q_a^{(m+1)}$, and similar for $b$. But this means that $T(Q_1)=Q_2$ and vice-versa. That is, translation acts trivially on the operators comprising the local copies of $G_{int}$, but nontrivially on the remaining global symmetry $G_m$.

To understand Eq.~\ref{eqn:0CellEquivalence}, consider inserting a charge of $Q_1$ (i.e. the $[1,0]$ element of $\H^1(G_m,\U1)$) on site $m$, where for concreteness we take $m$ to be even. This is implemented by acting with $Z_b$ on any site in the $m$th 1-cell, and thus is a charge of $Q^{(m)}_b$ on said 1-cell. If we want to do this in a translation-invariant way, then we must act with $T(Z^{(m)}_b)=Z^{(m+1)}_b$ on the $(m+1)$st 1-cell, that is, we must insert an excitation charged under $Q_2$ and $Q_b^{(m+1)}$ on the $(m+1)$ 1-cell. Since the excitation on the $(m+1)$st 1-cell is therefore the $[0,1]$ element of $\H^1(G_m,\U1)$ under the global symmetry, our global symmetry data on the $m$th 1-cell gets acted on by $T^\ast$ in order to preserve translation.

Observe that by acting with terms in the Hamiltonian Eq.~\ref{eqn:clusterH1} within the 1-cell, this charge may hop to site $\ell$ while charged under $Q^{(m)}_b$ (and, of course, $Q_1$). Once the charge reaches site $\ell$, the first term in Eq.~\ref{eqn:exactlySolubleCouple} allows the charge to hop into the $(m+1)$ 1-cell, but the state becomes charged under $Q^{(m+1)}_a$ and $Q_1$. We now have $[1,0] \in \H^1(G_m,\U1)$ on 1-cell $(m+1)$. By hopping, the data $[1,0]$ on site $(m)$ can become the data $[1,0]$ on site $(m+1)$; by doing this hopping in a translation-invariant way, we see that the data $[0,1]$ is now associated with the $m$th 1-cell. Hence, the data $[1,0]$ and $[0,1]$ placed on the $m$th 1-cell are adiabatically connected and should be viewed as equivalent.

The key lesson here is that when symmetry charges move, i.e. by hopping, their cohomology data under the \textit{modulated} symmetry is unchanged, but their cohomology data under the \textit{local} copies of the $G_{int}$ is acted upon by $T^\ast$. But when symmetry charges are \textit{acted upon by translation}, the situation is reversed; their cohomology data under the \textit{modulated} symmetry is acted upon by $T^\ast$, but their cohomology data under the \textit{local} symmetry is unchanged.

The second fact is why it naively appears that $\H^1(G_{int},\U1)$ classifies the weak SPT data in (1+1)D; the first fact is why we need to mod out by the equivalence~\ref{eqn:0CellEquivalence}.

\bibliographystyle{unsrt}
\bibliography{references}

\end{document}